\documentclass[aps,amsmath,superscriptaddress,amssymb,twocolumn,longbibliography,floatfix]{revtex4-2}

\usepackage{mathrsfs} 
\usepackage{graphicx,color} 
\usepackage{wasysym}
\usepackage{mathtools}
\usepackage{bigints}
\usepackage{bm}
\usepackage{threeparttable}
\usepackage[normalem]{ulem}
\usepackage{physics}
\usepackage[colorlinks=true]{hyperref}
\hypersetup{
  colorlinks=true,
  linkcolor=blue,
  filecolor=magenta,
  citecolor=blue,
  urlcolor=blue,
}

\newcommand{\red}[1]{\textcolor{black}{#1}}

\newcommand{\ovl}{\overline}

\newcommand{\mcat}{m_{\rm cat}}
\newcommand{\pin}{p_u^{\rm (in)}}
\newcommand{\pout}{p_v^{\rm (out)}}
\newcommand{\pcat}{p_w^{\rm (cat)}}
\newcommand{\mP}[1]{\mathcal{P}\qty[#1]}
\newcommand{\refbf}[1]{\textbf{\ref{#1}}}

\begin{document}

\title{
Starvation suppression in dense scale-free metabolic networks: Dynamical mean-field analysis of catalytic reaction networks
}

 \author{Kota Mitsumoto}
 \email{kmitsumoto@g.ecc.u-tokyo.ac.jp}
 \affiliation{Graduate School of Arts and Sciences, University of Tokyo, Komaba 3-8-1, Meguro-ku, Tokyo 153-8902, Japan}
 \author{Shuji Ishihara}%
 \email{csishihara@g.ecc.u-tokyo.ac.jp}
 \affiliation{Graduate School of Arts and Sciences, University of Tokyo, Komaba 3-8-1, Meguro-ku, Tokyo 153-8902, Japan}
 \affiliation{Universal Biology Institute, University of Tokyo, Komaba 3-8-1, Meguro-ku, Tokyo 153-8902, Japan}

\date{\today}

\begin{abstract}
Cellular metabolic networks exhibit scale-free topologies with power-law degree distributions across diverse organisms. Although such topologies are often linked to mutational robustness and evolutionary advantage, their role in metabolic dynamics remains unclear. Using dynamical mean-field theory, we derive an exact solution for an intracellular catalytic reaction model on dense random networks with arbitrary degree distributions. We show that the metabolic-starvation transition observed under nutrient-poor conditions for homogeneous degree distributions disappears when the out-degree distribution is scale-free. 
\red{We also show a power-law in-degree distribution of the underlying catalytic reaction network gives rise to a power-law distribution of biomolecular abundances with the same exponent. Large-scale numerical simulations validate these predictions. Our results provide a theoretical framework linking network topology and metabolic dynamics.}
\end{abstract}

\maketitle
%



Living systems sustain diverse vital functions through their intrinsic complexity~\cite{kauffman1993the,kaneko2006life,kaneko2025universal}. Even a single cell maintains metabolism through a rich mixture of chemical species interacting within extremely complex reaction networks~\cite{ramirezgaona2016YMDB, haug2020metaboLights, wishart2022HMDB, kim2025PubChem, kanehisa2025KEGG}. These metabolic networks vary substantially across species and cellular functions, leading to highly diverse detailed structures. Despite this diversity, metabolic networks commonly possess a scale-free topology~\cite{albert2002statistical, barabasi2004network, barabasi2016network}. A scale-free topology is defined by a power-law degree distribution, which gives rise to a small number of highly connected hubs~\cite{jeong2000the,jeong2001lethality,wagner2001the,ravasz2002hierarchical,Tong2004,li2004map,nacher2005two}.
\red{Scale-free structures have also been reported in many other complex systems, including social~\cite{ebel2002scalefree,ferrara2012community}, technological~\cite{lin2013transportation, nesti2020emergence}, and information networks~\cite{barabasi2000scale, pastorsatorras2001dynamical}. Although a recent study has shown that not all reported examples are scale-free~\cite{broido2019scale}, these findings do not invalidate the characterization of metabolic networks as scale-free. Scale-free topologies have been argued to confer robustness against the random removal of nodes or edges~\cite{albert2000error, cohen2000resilience, cooper2006effect, cohen2010complex, artime2024robustness}. In metabolic networks, such robustness may help preserve metabolic function despite mutations, and the resulting evolutionary advantage may partly explain the widespread occurrence of scale-free topology in these networks~\cite{Oikonomou2006,greenbury2010effect}.}

\red{Apart from} evolutionary advantages, scale-free networks are known to influence the collective behavior~\cite{newman2010networks, barabasi2016network, albert2002statistical, barzel2013universality, dorogovtsev2008critical}.
One important consequence is the elimination of phase transitions, as seen in the ferromagnetic Ising model~\cite{aleksiejuk2002ferromagnetic} and the epidemic spreading model~\cite{pastor-satorras2001epidemic}. In these systems, hubs effectively couple a large fraction of nodes and thereby promote global ordering. Metabolic systems likewise exhibit qualitative changes in cellular states, including transitions between living and non-living states~\cite{newton2024cell, himeoka2024theoretical} and cell differentiation~\cite{furusawa2012dynamical, lanza2014essentials}. 
Scale-free topology may therefore influence not only the mutational robustness of metabolic systems but also their dynamical phase behavior. 
However, it remains unclear how the scale-free topology of metabolic networks influences these transitions, including whether it can help cells avoid death or starvation.

\begin{figure*}[t]
\centering
\includegraphics[width=170mm]{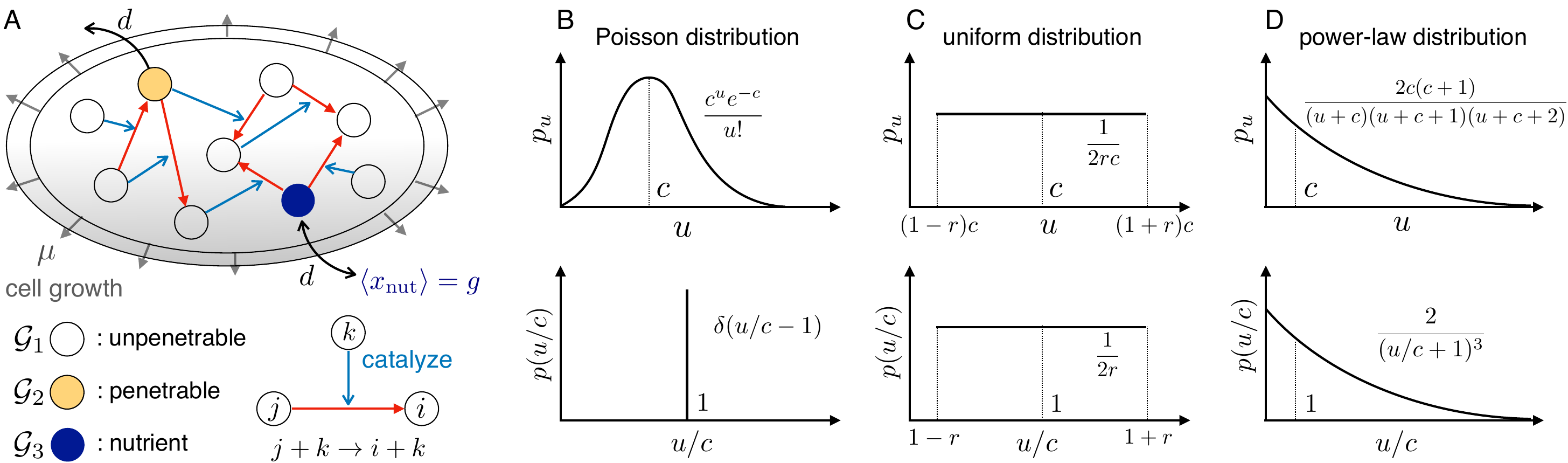}
\caption{
(A) Schematic illustration of a model of a cell incorporating catalytic reactions, nutrient uptake, and cell growth. The catalytic reaction $j+k\to i+k$ represents the conversion of the chemical species $j$ into the $i$ catalyzed by the $k$. The model consists of three groups of chemical species: unpenetrable metabolic products, penetrable metabolic products, and nutrients, denoted by $\mathcal{G}_1, \mathcal{G}_2$ and $\mathcal{G}_3$, respectively. Nutrients are catalytically inactive. Penetrable metabolic products and nutrients can pass through the cell membrane with the permeability $d$. Their equilibrium abundances in the external environment are 0 and $g$, respectively. The cell grows at a rate $\mu$, which is also a dynamical variable. Schematic illustrations of (B) Poisson, (C) uniform, and (D) power-law ($\beta=3$) distributions with average connectivity $c$. Distributions of sparse and dense regimes are displayed at the top and bottom, respectively.
}
\label{fig:schematic}
\end{figure*}

To understand how metabolic dynamics depends on network topology, one needs to compare systems with different topologies under the same environmental conditions.
A useful strategy is therefore to employ a minimal model that captures essential mechanisms while abstracting away microscopic details.
A simple catalytic reaction network model incorporating cell growth and nutrient uptake was introduced by Furusawa and Kaneko (see Fig.~\ref{fig:schematic}A)~\cite{furusawa2001theory, furusawa2003zipf, furusawa2006evolutionary, furusawa2012adaptation}. 
Using evolutionary simulations under selection for high cellular growth rates, they showed that this model gives rise to both a power-law distribution of biomolecular abundances and a power-law degree distribution of the reaction network, i.e., scale-free topology. These results provide a possible explanation for the power-law distributions of biomolecular abundances observed in real cells, including those of mRNAs and metabolites~\cite{furusawa2003zipf, ueda2004universality, lu2009investigation, sato2018power, lazzardi2023emergent}. However, the contribution of the scale-free topology to the high cellular growth rate and the biomolecular abundance distribution, as well as the transition of the metabolic states, remains unclear, partly due to the lack of exact analytical results beyond numerical simulations. To address this issue, we analyze random network models in the limits of large system size and dense connectivity. In these limits, an exact analytical treatment becomes possible using dynamical mean-field theory (DMFT)~\cite{martin1973statistical,janssen1976on,dedominicis1978dynamics,sompolinsky1982relaxational,cugliandolo1993analytical,castellani2005spin,galla2024generating}, which is based on a generating-functional formalism for dynamical paths and enables ensemble averaging over random network topologies. Furthermore, recent technical developments have extended this framework to networks with arbitrary degree distributions~\cite{mimura2009parallel,poley2025interaction,metz2025dynamical}, thereby allowing systematic comparisons among networks with different degree distributions, including scale-free networks, as displayed in Fig.~\ref{fig:schematic}B.

In this paper, we analyze a densely connected catalytic reaction network model using DMFT. Since catalytic reactions are directional, we extend the DMFT framework to directed networks with arbitrary in- and out-degree distributions (See {\it SI Appendix}, Text S1). We derive exact effective dynamical equations conditioned on in- and out-degrees and obtain their fixed points. We show that the model exhibits three states: metabolic, overnutrition, and starvation states. The transition between metabolic and overnutrition states occurs under high nutrient supply via a transcritical bifurcation, and the phase boundary is independent of both in- and out-degree distribution. On the other hand, the boundary of the metabolic-starvation transition, which occurs under low nutrient supply, is highly sensitive to the out-degree distribution. In particular, the transition to the starvation state disappears for power-law out-degree distributions, i.e., scale-free networks. Furthermore, we find that the effect of connectivity heterogeneity reverses depending on nutrient availability: under nutrient-rich conditions, larger heterogeneity suppresses the growth rate, whereas under nutrient-poor conditions it enhances growth and prevents the starvation state. This implies that scale-free networks can provide an advantage in nutrient-poor environments. In addition, we \red{examine} the validity of the dense-limit results by comparing with numerical simulations on Erd\H{o}s-R\'{e}nyi (ER) networks~\cite{erdos1959random, newman2010networks, barabasi2016network} and one of scale-free networks called directed Barab\'{a}si-Albert (dBA) networks~\cite{dorogovtsev2000structure, barabasi2016network} with finite connectivity $c$ 
\red{, defined as the average number of edges per node (see {\it Materials and Methods}). From this comparison, we find} quantitative agreements down to $c\approx 10$. \red{For smaller $c$, however, network sparsity itself also helps suppress the starvation state, making it difficult to distinguish the contribution of the power-law out-degree distribution from that of sparsity. We therefore discuss the range of $c$ in which the effect of the power-law out-degree distribution remains dominant over the effect of sparsity.} 
We also show that a power-law in-degree distribution\red{, rather than a power-law out-degree distribution,} yields a power-law tail in the abundance distribution of biomolecules. This prediction is also confirmed by numerical simulations on dBA networks.
\red{Finally, we discuss the implications of the present study and its limitations in terms of modeling and analytical methods.}


\section*{Model}
We adopt the model used in~\cite{furusawa2003zipf,furusawa2006evolutionary,furusawa2012adaptation}, in which a cell is represented as a membrane-bound compartment containing $N$ chemical species that interact through catalytic reactions on a random network, as illustrated in Fig.~\ref{fig:schematic}A. The abundance of each chemical species is denoted by $x_i(t)$ for $i=1,2,...,N$. The catalytic reactions in this model take the simple form $j+k\to i+k$, representing the conversion of the chemical species $j$ into $i$, catalyzed by $k$. In this reaction, the production rate of $x_i$ and consumption rate of $x_j$ are both given by $x_j x_k$, thereby the total abundance $x_i + x_j$ is conserved. The chemical species are divided into three groups: unpenetrable metabolic products, penetrable metabolic products, and nutrients, denoted by $\mathcal{G}_1, \mathcal{G}_2$ and $\mathcal{G}_3$, respectively. The fraction of chemical species in $\mathcal{G}_n$ is denoted by $\alpha_n$, satisfying $\alpha_1 + \alpha_2 + \alpha_3 = 1$. Nutrients are assumed to be catalytically inactive. \red{The simplification of the chemical reaction and the classification into $\mathcal{G}_1$, $\mathcal{G}_2$, and $\mathcal{G}_3$ follows the original model in Refs~\cite{furusawa2003zipf,furusawa2006evolutionary,furusawa2012adaptation}. We note that each chemical species in these groups should be interpreted as a coarse-grained representation of biomolecules and do not correspond one-to-one with proteins, mRNA or metabolites.} 

Penetrable metabolic products and nutrients can pass through the membrane and exchange with the external environment at rates $-dx_i$ and $d(g-x_i)$, respectively, where $d$ is the membrane permeability, and $g$ is the equilibrium abundance of nutrients in the environment. For convenience, we introduce parameters $(d_i, g_i)$, defined as $(0,0)$ for $i \in \mathcal{G}_1$, $(d,0)$ for $i \in \mathcal{G}_2$, and $(d,g)$ for $i \in \mathcal{G}_3$. The cell grows by taking up nutrients from the environment; accordingly, the abundances of chemical species are uniformly diluted by $-\mu x_i$, where $\mu = \frac{1}{N}\sum_{i =1}^N d_i(g_i-x_i)$ is the cell growth rate. 
\red{Here and below, the abundance $x_i$, time $t$, permeability $d$, equilibrium nutrient abundance $g$, and growth rate $\mu$ are dimensionless; the nondimensionalization procedure is described in {\it Materials and Methods}. This nondimensionalization assumes that the cell-volume growth is determined by the net increase in the total amount of chemical species, which leads to the above definition of the cell growth rate.}

Taken together, the dynamics of $x_i(t)$ is governed by the following equations:
\begin{align}
\nonumber
\dot{x}_i(t) &= \frac{1}{c}\sum_{j =1}^N \sum_{k \notin \mathcal{G}_3} C_{ij}^k x_j(t) x_k(t) -  \frac{1}{c}\sum_{j =1}^N \sum_{k \notin \mathcal{G}_3}C_{ji}^k x_i(t) x_k(t) \\
 &+ d_i(g_i-x_i(t)) - \mu(t) x_i(t).
 \label{eq:dynamical1}
\end{align}
The first and the second terms represent the catalytic reactions incoming to and outgoing from the $i$-th chemical species, respectively. The tensor $\{ C_{ij}^k \}$ encodes the topology of catalytic reaction network, where $C_{ij}^k$ is equal to $1$ if  reaction $j+k\to i+k$ exists, and is equal to $0$ otherwise. $c$ is the mean connectivity of the network, i.e., $\frac{1}{N}\sum_{i,j,k}C_{ij}^k = c$. \red{The factor $1/c$ normalizes the catalytic reaction terms so that the reaction flux into each species remains $\mathcal{O}(1)$ when the mean connectivity $c$ is large. This normalization enables us to compare networks with different $c$ under the scaled reaction rate and to calculate the dense-connectivity limit exactly using DMFT (see SI Appendix, Text S1).} The third and fourth terms represent the interaction with the external environment and dilution due to cell growth, respectively. By taking the sum of Eq.~\refbf{eq:dynamical1} over $i$, one finds $\sum_{i=1}^N \dot{x}_i(t) = \mu(t) (N-\sum_{i=1}^N x_i(t))$; hence, positive $\mu(t)$ enforces a constraint on the long-time behavior of the average abundance of chemical species, leading to
\begin{align}
    \lim_{t\to \infty}\frac{1}{N}\sum_{i=1}^N x_i(t) = 1.
    \label{eq:constraint1}
\end{align}
If $\mu<0$, the cell shrinks and eventually dies. In the model, we assume constant catalytic reaction rates following previous studies~\cite{furusawa2003zipf, furusawa2006evolutionary}. We do not explicitly account for cell division, as its inclusion does not affect the conclusions of the analysis presented below.


To derive the effective dynamical equations on random networks, we adopt the configuration model~\cite{bollobs1980a, chung2002connected, newman2010networks, poley2025interaction}, which allows the construction of random networks with arbitrary distributions of in-degrees $\pin$ and out-degrees $\pout$, both having mean $c$, and catalytic degrees $\pcat$ with mean $c/(\alpha_1+\alpha_2)$. \red{Here, the in-degree and out-degree of a node count the numbers of incoming and outgoing reaction edges, respectively. The catalytic degree, on the other hand, counts the numbers edges catalyzed by the node, as illustrated in Fig.~\ref{fig:schematic}A.} To construct a network, we first independently draw the degrees $u_i, v_i$, and $w_i$ for each node from distributions $\pin, \pout$, and $\pcat$, respectively. For nodes belonging to $\mathcal{G}_3$, we set $w_i = 0$ because nutrients are catalytically inactive. The connection $C_{ij}^k$ is then set to $1$ with probability $u_iv_jw_k/(Nc)^2$ and to $0$ otherwise. For sufficiently large $N$, this construction yields networks with the prescribed degree distributions. Below, we consider three kinds of distributions: a Poisson distribution (corresponding to the ER network), a uniform distribution defined on the interval $[(1-r)c, (1+r)c]$~($0<r\le 1$), and a power-law distribution with exponent $\beta~(2 < \beta)$, which are displayed in Figs.~\ref{fig:schematic}B, C and D, respectively. Networks with uniform or power-law degree distributions retain structural heterogeneity even in the dense limit, $c \to \infty$, whereas the ER network approaches a regular (homogeneous) random network.

\section*{Results}


\subsection*{Effective dynamical equations and fixed points}

In the thermodynamic limit, $N\to\infty$, and dense connectivity, $c \gg 1$, we analyze the dynamics governed by Eq.~\refbf{eq:dynamical1} using DMFT. By averaging over the network topology for given degree distributions, the effective dynamical equations for the typical abundances of chemical species belonging to $\mathcal{G}_n$ with in-degree $u$ and out-degree $v$ are obtained exactly. The resulting dynamics is independent of the catalytic degree $w$ because the catalyzed chemical species do not provide feedback through the catalytic activity.  The derivation of the DMFT equations is presented in {\it SI Appendix}, Text S1. The effective dynamical equations are given by
\begin{align}
\nonumber
\dot{x}_{nuv}(t) &= \frac{u}{c}m_{\rm in}(t) \mcat(t) - \frac{v}{c} \mcat(t) x_{nuv}(t) \\
&+ d_n(g_n-x_{nuv}(t)) - \mu(t) x_{nuv}(t),
\label{eq:dynamical2}
\end{align}
where $m_{\rm in}(t) = \sum_{n,u,v} \frac{v}{c} \alpha_n \pin\pout  x_{nuv}(t)$ and $\mcat(t) = (\alpha_1 m_1(t) + \alpha_2 m_2(t))/(\alpha_1 + \alpha_2)$. $m_n(t) = \sum_{u,v} \pin\pout x_{nuv}(t)$ represents the average abundance of chemical species belonging to $\mathcal{G}_n$. 
Hence, $\mcat$ is regarded as the average abundance of catalytic chemical species. The growth rate is given by $\mu(t) = d(\alpha_3 g -\alpha_2 m_2(t) -\alpha_3 m_3(t))$. By taking the average of Eq.~\refbf{eq:dynamical2} over $u, v$, and $n$, and using the definition of $m_{\rm in}(t)$, one finds that the average abundance of all chemical species converges to
\begin{align}
\lim_{t \to \infty}(\alpha_1 m_1(t) + \alpha_2 m_2(t) + \alpha_3 m_3(t)) = 1
\label{eq:constraint2}
\end{align}
if $\mu(t)>0$, corresponding to Eq.~\refbf{eq:constraint1}. 


At the fixed point, the following relation holds:
\begin{align}
    x_{nuv}^* = \frac{d_n g_n + \frac{u}{c}m_{\rm in}^* \mcat^*}{d_n + \mu^* + \frac{v}{c}\mcat^*},
    \label{eq:fixed}
\end{align}
where $A^* = \lim_{t\to\infty}A(t)$. One finds that the chemical abundance increases with increasing in-degree $u$ and decreases with increasing out-degree $v$. 
By averaging $x_{nuv}^*$ over $u$ and $v$, we obtain 
\begin{align}
m_n^* = (d_n g_n + m_{\rm in}^* \mcat^*)X(\mu^*+d_n, \mcat^*),
\label{eq:m_n}
\end{align}
where $X(a,b) = \sum_{v}\pout/(a+\frac{v}{c}b)$. 
Notably, $m_n^*$ does not depend on the in-degree distribution; hence, macroscopic quantities such as $\mu^*$ and $\mcat^*$ are independent of the in-degree distribution. 
Using the definitions of $\mu^*$ and $\mcat^*$, together with Eq.~\refbf{eq:constraint2}, we obtain a closed set of equations for $\mu^*, m_{\rm in}^*,$ and $\mcat^*$:
\begin{align}
m_{\rm in}^* \mcat^* &=  \frac{(\alpha_1 + \alpha_2)\mcat^*}{\alpha_1 X(\mu^*, \mcat^*) + \alpha_2 X(\mu^*+d, \mcat^*)}, \label{eq:set_equation1} \\
m_{\rm in}^* \mcat^* &= \frac{d\alpha_3 g(1 -dX(\mu^*+d, \mcat^*))  - \mu^*}{d(1-\alpha_1)X(\mu^*+d, \mcat^*)}, \label{eq:set_equation2}\\
m_{\rm in}^* \mcat^* &= \frac{1-d\alpha_3 gX(\mu^*+d, \mcat^*)}{\alpha_1X(\mu^*, \mcat^*) + (1-\alpha_1)X(\mu^*+d, \mcat^*)}.\label{eq:set_equation3}
\end{align}

These equations admit two distinct solutions, corresponding to an overnutrition state and a metabolic state. In the overnutrition state, the cell is occupied solely by nutrients, i.e., $\mcat^* = 0$, immediately obtained from Eq.~\refbf{eq:set_equation1}. Since nutrients are catalytically inactive, no catalytic reactions occur inside the cell in this state. Accordingly, $x_{1uv}^*$ and $x_{2uv}^*$ vanish, while $x_{3uv}^*=1/\alpha_3$ for any $u$ and $v$. The growth rate is then given by $\mu^* = d(\alpha_3 g-1)$, which is independent of both the in- and out-degree distributions. 
\red{The positive growth rate in the absence of catalytic reactions results from the simplified assumption of a constant external nutrient supply and the neglect of mechanical constraints such as membrane elasticity (see {\it Materials and Methods}).} By contrast, in the metabolic state, all chemical species have finite abundances, and catalytic reactions are sustained within the cell. For Poisson and uniform degree distributions, the cell growth rates of both solutions become negative if the nutrient supply is too small. Then, the convergence condition of Eq.~\refbf{eq:constraint2} is violated, indicating that the cell shrinks and eventually dies. In this study, we define the state located in this parameter region as a starvation state.


\subsection*{Phase diagram}

\begin{figure}[t]
\centering
\includegraphics[width=85mm]{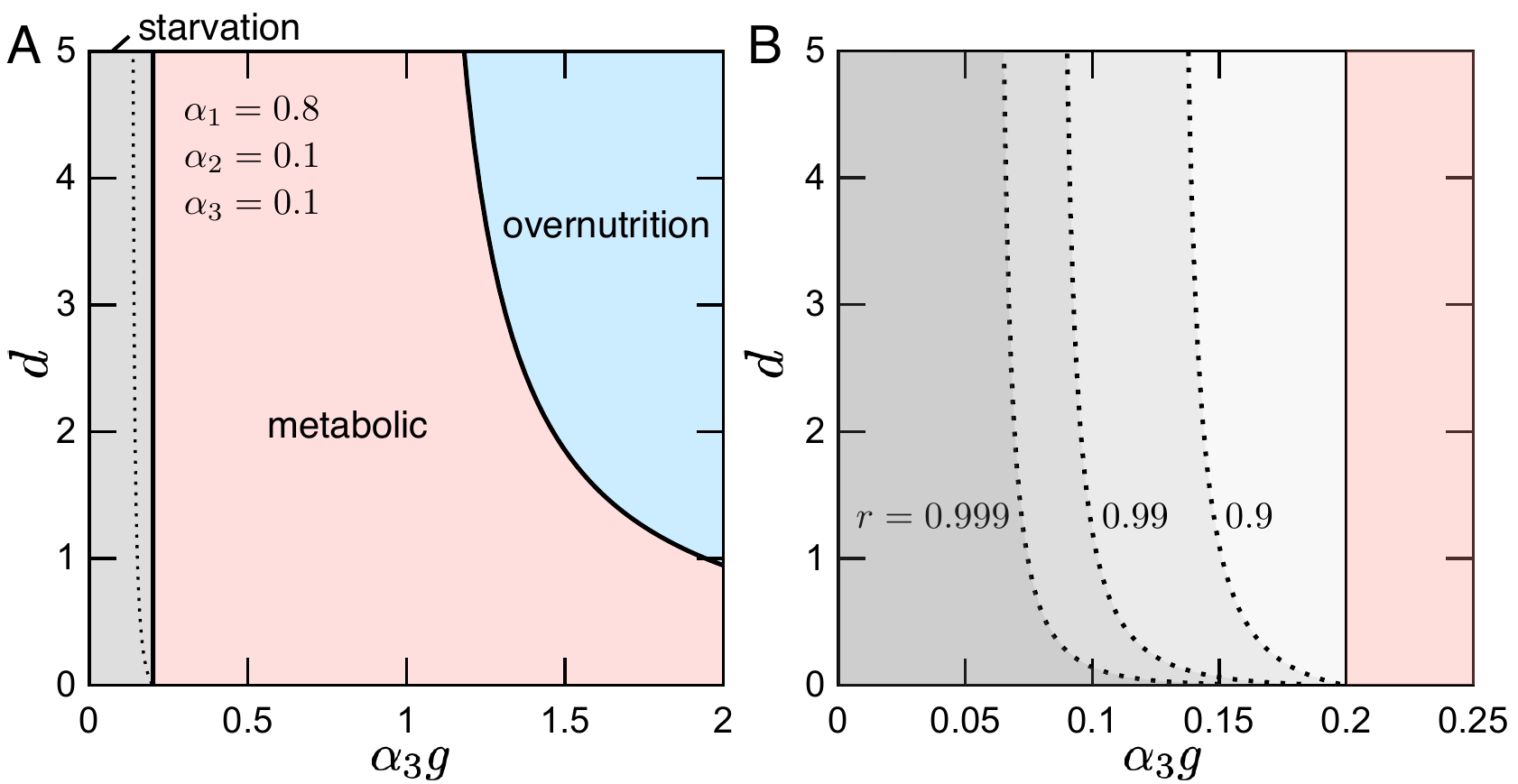}
\caption{
(A) Phase diagram in the dense-network limit for the parameters $(\alpha_1,\alpha_2,\alpha_3) = (0.8,0.1,0.1)$. The solid curve represents the boundary $d_{\rm c}$ between metabolic and overnutrition states, given in Eq.~\refbf{eq:critical}. The solid vertical line, $\alpha_3g = 1-\alpha_1$, and dashed curve represent the boundary between metabolic and starvation states for Poisson and uniform ($r=0.9$) distributions, respectively. (B) Phase diagram for low $\alpha_3 g$ with $(\alpha_1,\alpha_2,\alpha_3) = (0.8,0.1,0.1)$. The phase boundaries between the metabolic and starvation states for $r = 0.9, 0.99$, and $0.999$ are shown.
}
\label{fig:phase}
\end{figure}

We first present in Fig.~\ref{fig:phase}A the phase diagram in the $(\alpha_3 g, d)$ plane for $(\alpha_1, \alpha_2, \alpha_3) = (0.8, 0.1, 0.1)$. For large $\alpha_3 g$, the transition between metabolic and overnutrition states occurs. The phase boundary is given by 
\begin{align}
d_{\rm c} = \frac{\alpha_3 g - \alpha_2/(\alpha_1+\alpha_2)}{\alpha_3 g (\alpha_3 g-1)},
\label{eq:critical}
\end{align}
which is obtained from a linear stability analysis of the overnutrition solution~\cite{opper1992phase, roy2019numerical, galla2024generating} (see {\it SI Appendix}, Text S2 for derivation). Notably, the onset of instability is independent of the degree distribution. The system tends to be in an overnutrition state at higher values of $\alpha_3 g$ and $d$, where nutrient supply dominates and effectively suppresses catalytic reactions. Moreover, the phase boundary shifts downward with an increasing fraction of penetrable metabolic products $\alpha_2$, since they are subject to a loss of abundance through interaction with the external environment. 


In contrast to the metabolic-overnutrition transition, the boundary of the metabolic-starvation transition at low $\alpha_3g$ depends sensitively on the out-degree distributions, as shown in Fig.~\ref{fig:phase}B. For the Poisson distribution, the boundary is $\alpha_3g = 1 -\alpha_1$ at which the cell growth rate becomes negative (see {\it SI Appendix}, Text S3), whereas for uniform and power-law distributions, the metabolic state persists even for $\alpha_3g < 1 -\alpha_1$. In particular, the starvation phase disappears for the power-law distribution. For the uniform distribution, the phase boundary shifts downward with increasing the heterogeneity parameter $r$, and the starvation phase disappears at $r=1$. Notice that even for values very close to $r=1$, such as $r=0.999$, the phase boundary remains robustly present, indicating $r=1$ is singular.  We will present the behaviors of physical quantities at low $\alpha_3g$ and the reason for the disappearance of the starvation state in detail \red{later}.


\subsection*{Transition between metabolic and overnutrition states}

\begin{figure}[t]
\centering
\includegraphics[width=85mm]{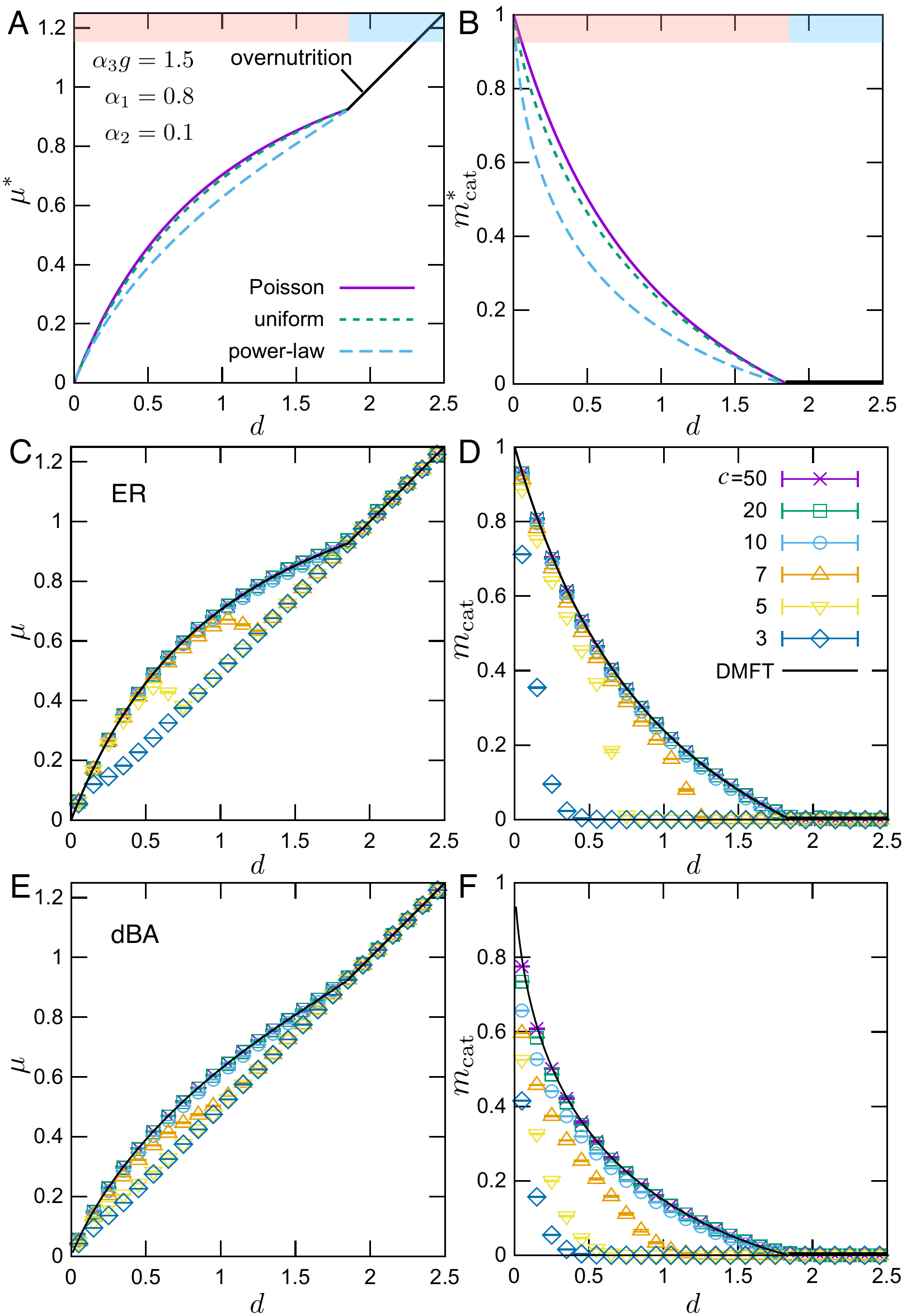}
\caption{
Dependence on permeability $d$ of (A) the growth rate $\mu^*$ and (B) the average abundance of catalytic chemical species $\mcat^*$ in steady state, obtained from the DMFT. In the metabolic state $d<d_{\rm c}\approx 1.85$, different colors represent different network topologies: results for Poisson, uniform ($r=1$), and power-law degree distributions (exponent $\beta = 3$) are shown from top to bottom. The black line for $d>d_{\rm c}$ represents the non-metabolic state. Dependence on $d$ of (C) $\mu(t)$ and (D) $\mcat(t)$ at $t=100$ for the ER networks with system size $N=10^5$ and finite connectivities $c = 50, 20, 10, 7, 5,$ and 3. Dependence on $d$ of (E) $\mu(t)$ and (F) $\mcat(t)$ at $t=100$ for the dBA networks with system size $N=10^5$ and finite connectivities $c = 50, 20, 10, 7, 5,$ and 3. The dense-limit result from the DMFT is also shown for comparison. $\alpha_3g=1.5$, $\alpha_1=0.8$, and $\alpha_2=0.1$ are used in all panels. The simulation data are averaged over 100 network realizations.
}
\label{fig:mu}
\end{figure}

We now present the behavior of the physical quantities in the metabolic-overnutrition transition that occurs at high nutrient supply. Since the metabolic solution depends on the out-degree distribution, we analyze it for Poisson, uniform, and power-law distributions, which yield homogeneous, moderately heterogeneous, and strongly heterogeneous networks in the dense limit, respectively. For the Poisson degree distribution, the metabolic solution can be derived analytically (see {\it SI Appendix}, Text S3 for details). By contrast, for the uniform and power-law degree distributions, the solution is not available analytically and is therefore obtained numerically using the Levenberg-Marquardt algorithm~\cite{press2007numerical}. 

Figures~\ref{fig:mu}A and B show the growth rate $\mu^*$ and the average abundance of the catalytic chemical species $\mcat^*$, respectively, as functions of $d$, for $\alpha_3g=1.5$, $\alpha_1=0.8$, and $\alpha_2=0.1$. 
The value of $\mu^*$ for $d < d_{\rm c}$ is largest in the Poisson case and smallest in the power-law case, suggesting that increasing network heterogeneity suppresses the growth rate for high $\alpha_3 g$. 
As $d$ approaches $d_{\rm c}$, $\mu^*$ converges to the same value, and the system undergoes a transition to the non-metabolic state via a transcritical bifurcation at $d = d_{\rm c}$. Likewise, $\mcat^*$ decays to zero continuously, as shown in Fig.~\ref{fig:mu}B.


To examine the range of validity of our theory with respect to network connectivity, $c$, we perform numerical simulations of Eq.~\refbf{eq:dynamical1} on ER and dBA networks with finite connectivity and $N=1.0\times10^5$ (see {\it Materials and Methods} for details). Here, the dBA networks have the power-law out-degree distribution with exponent $\beta=3$ and homogeneous in-degrees. In Figs.~\ref{fig:mu}C and D, we show $\mu(t)$ and $\mcat(t)$ evaluated at $t=100$, which is sufficiently long for the system to reach the fixed point, except in the vicinity of the transition point for small connectivities (see {\it SI Appendix}, Figs.~S1A and B). The data for $c=50, 20, 10$ collapse onto the DMFT result, indicating that the network with $c \gtrsim10$ is sufficiently dense. For $c < 10$,  the data agree with the DMFT result at small $d$ but exhibit a sudden shift to the overnutrition state below the DMFT transition point. Thus, network sparsity destabilizes the metabolic state. Similar to ER networks, the data for $c=50, 20, 10$ on the dBA networks exhibit a good agreement with the DMFT result for the power-law distribution, as shown in Figs.~\ref{fig:mu}E and F. For $c<10$, $\mu$ gradually shifts downward from the DMFT result with increasing $d$. This is also due to the destabilization of the metabolic state.


\subsection*{Transition between metabolic and starvation states}
\label{sec:starved}

\begin{figure}[t]
\centering
\includegraphics[width=85mm]{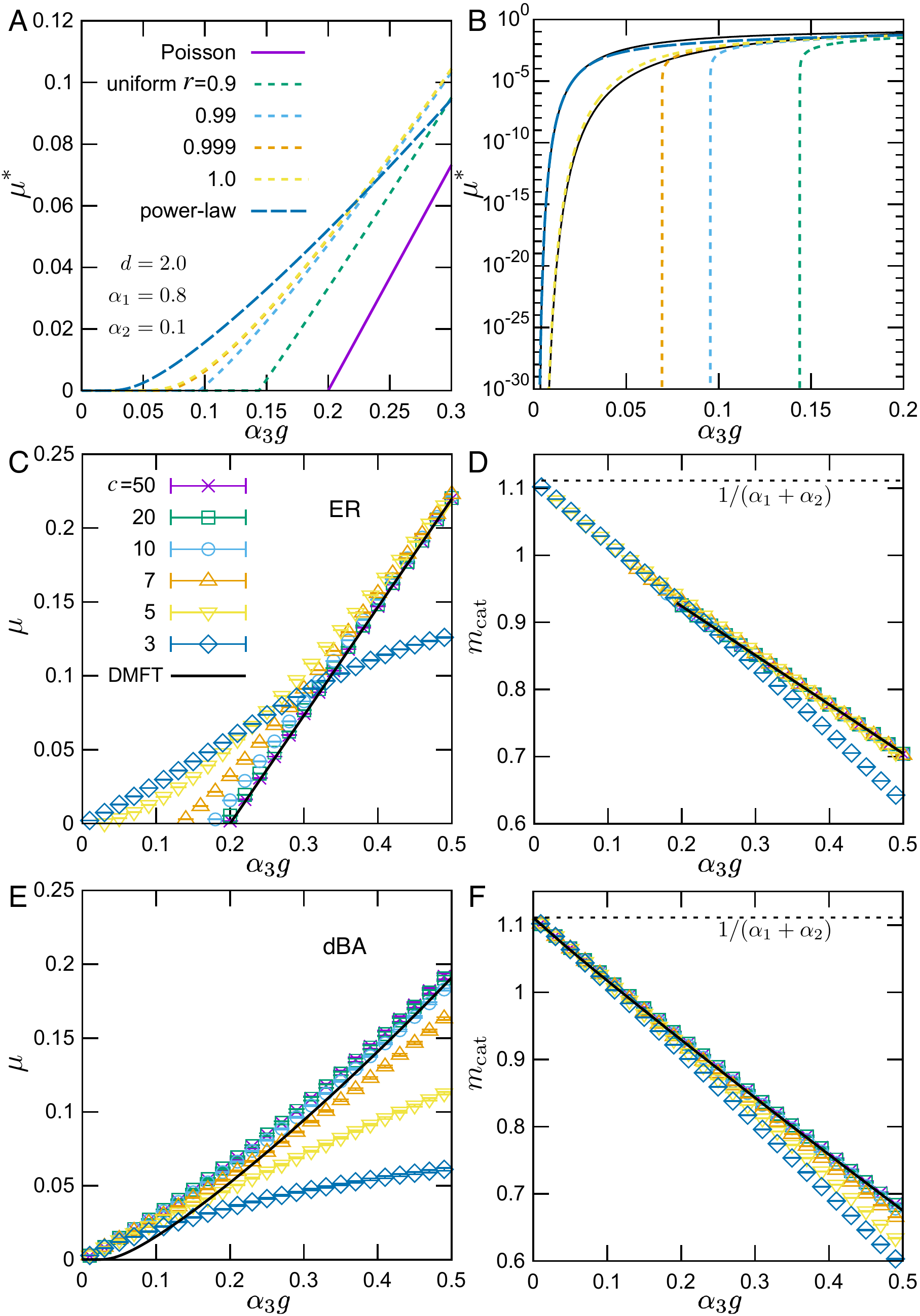}
\caption{
(A) Dependence on nutrient supply $\alpha_3g$ of the growth rate $\mu^*$ obtained from the DMFT and (B) its semi-log plot. Different colors represent different network topologies: results for Poisson, uniform ($r=0.9,0.99,0.999,1$), and power-law degree distributions (exponent $\beta = 3$) are shown. The black solid curves in panel (B) represent asymptotic solutions for $\alpha_3g \ll 1$. Dependence on $\alpha_3g$ of (C) $\mu(t)$ and (D) $\mcat(t)$ at $t=100$ for the ER networks with system size $N=10^5$ and finite connectivities $c = 50, 20, 10, 7, 5,$ and 3. Dependence on $\alpha_3g$ of (E) $\mu(t)$ and (F) $\mcat(t)$ at $t=100$ for the dBA networks with system size $N=10^5$ and finite connectivities $c = 50, 20, 10, 7, 5,$ and 3. The dense-limit result from the DMFT is also shown for comparison. The dotted horizontal lines in panels (D) and (F) represent the maximum value of $m_{\rm cat}$, $1/(\alpha_1 + \alpha_2)$. $d=2.0$, $\alpha_1=0.8$, and $\alpha_2=0.1$ are used in all panels. The simulation data are averaged over 100 network realizations.
}
\label{fig:starved}
\end{figure}

Next, we present the DMFT results for low nutrient supply. Figure~\ref{fig:starved}A displays the $\alpha_3 g$ dependence of the growth rate $\mu^*$ for Poisson, uniform ($r=0.9,0.99,0.999,1$), and power-law ($\beta=3$) out-degree distributions with $d=2.0$, $\alpha_1=0.8$, and $\alpha_2 = 0.1$. One finds that the value of $\mu^*$ is smallest in the Poisson case and largest in the power-law case for low $\alpha_3 g$, suggesting that a more heterogeneous network yields a higher growth rate, in contrast to the behavior at high $\alpha_3 g$. Indeed, for the uniform distribution, $\mu^*$ increases as the heterogeneity parameter $r$ increases. Furthermore, network heterogeneity makes cells more resilient to nutrient starvation. While $\mu^*$ becomes zero and exhibits a metabolic-starvation transition at $\alpha_3g = 1- \alpha_1$ for the Poisson distribution (see also {\it SI Appendix}, Text S3), the transition point shifts to lower $\alpha_3 g$ values for the uniform distribution with larger $r$, and disappears only when $r=1$, as shown in Fig.~\ref{fig:starved}B. The power-law distribution also does not exhibit such a transition. Mathematically, these disappearances of the metabolic-starvation transitions originate from the fact that $X(\mu^*, \mcat^*)$ appearing in $m_1^*$, given by \eqref{eq:m_n}, exhibits a logarithmic divergence as $\mu^* \to +0$ (see {\it SI Appendix}, Text S4 in detail). This divergence makes the mean abundance of unpenetrable metabolic products $m_1^*$, significantly larger than those of the other species, $m_2^*$ and $m_3^*$. More specifically, the logarithmic divergence of $X(\mu^*, \mcat^*)$ stems from the divergence of $x_{1uv}^*$, given by Eq.~\refbf{eq:fixed}, in the $v/c\to0$. This means that the existence of the starvation state depends on whether the network contains unpenetrable metabolic products with extremely small out-degrees, such that their outgoing reactions are negligible relative to their incoming reactions.
The asymptotic behavior of $\mu^*$ at $\alpha_3g \ll 1$ can be shown to take the exponential form $\mu^* \sim \exp(-C/\alpha_3 g)$, 
for both the uniform distribution with $r=1$ and the power-law distribution with $\beta=3$, where $C$ is a positive constant (see Eq.~{\bf S63} in {\it SI Appendix}, Text S4 and Fig.~\ref{fig:starved}B). 
Therefore, the cell growth rate $\mu^*$ becomes close to zero but remains positive for any $\alpha_3 g >0$.


In the same manner as in the high $\alpha_3g$ case, we also test the validity of the DMFT results with respect to network connectivity $c$ by performing numerical simulations of Eq.~\refbf{eq:dynamical1} for low $\alpha_3g$ on ER and dBA networks with finite connectivity and $N=1.0\times10^5$. In Figs.~\ref{fig:starved}C and D, we show $\mu(t)$ and $\mcat (t)$ evaluated at $t=100$. $\mu(t)$ for $c=50$ agrees well with the DMFT result but deviates upward for $c \le 20$. The transition point shifts to lower $\alpha_3g$ as $c$ decreases, and disappears for $c=3$, indicating that the network sparsity helps \red{avoid} starvation. On the other hand, $\mu(t)$ for $c=3$ becomes significantly smaller than the other data when $\alpha_3 g > 0.3$. In fact, in this parameter region, the dynamics have not yet converged at $t=100$ and appear to relax toward the overnutrition state (see {\it SI Appendix}, Figs.~S1C and D). Thus, while sparsity stabilizes metabolism at low $\alpha_3 g$, under nutrient-rich conditions the metabolic state becomes unstable, which is consistent with Fig.~\ref{fig:mu}C. $\mcat$ for all $c$ except $c = 3$ collapses well onto the DMFT result, but extends into the region $\alpha_3g < 1 -\alpha_1$. As $\alpha_3 g$ decreases, $\mcat$ approaches its maximum value, $1/(\alpha_1+\alpha_2)$, which corresponds to the limit $m_3=0$. Simulation results for low $\alpha_3g$ on the dBA networks do not show perfect agreement with the DMFT results, even when $c$ is large. As shown in Figs.~\ref{fig:starved}E and F, $\mu(t)$ for $c=50, 20$ is slightly higher than the DMFT result, while $\mcat(t)$ collapses onto the DMFT result. This discrepancy may be attributed to the small degree correlation between connected nodes in the dBA network~\cite{fotouhi2013degree}, which is not accounted for in our DMFT. Nevertheless, even with such an effect, the transition to the starvation state does not occur in scale-free networks.

\begin{figure}[t]
\centering
\includegraphics[width=85mm]{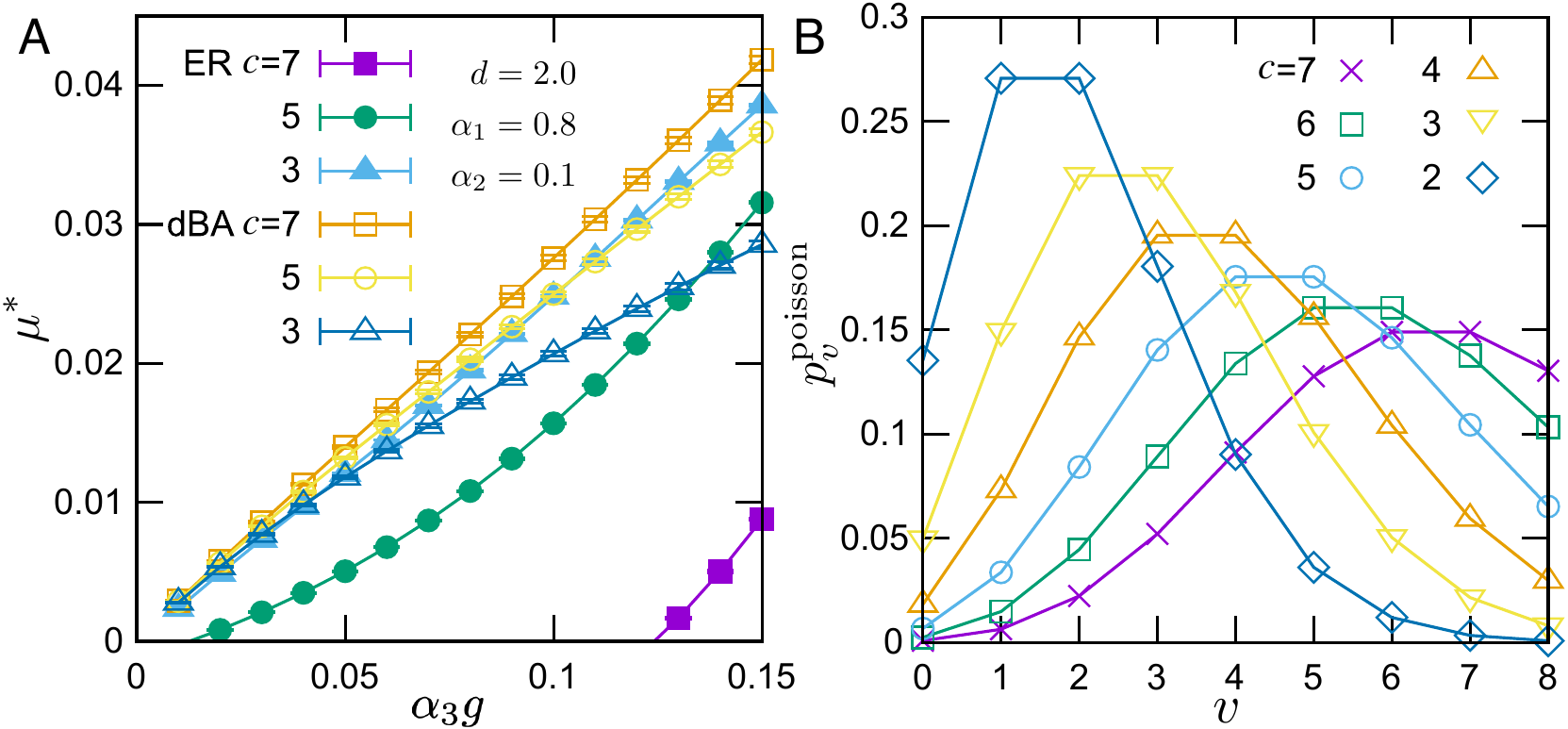}
\caption{
\red{(A) Direct comparison of the growth rate in ER and dBA networks with mean conductivities $c=7,5$, and 3 under nutrient-poor condition.  The parameters $d=2.0$, $\alpha_1=0.8$, and $\alpha_2=0.1$ are used. (B) Poisson distribution $p_v^{\rm poisson} = c^ve^{-c}/v!$ with mean $c=7,6,\ldots, 2$.}
}
\label{fig:compare}
\end{figure}

\red{These results suggest that both a power-law out-degree distribution and sparsity can help suppress the starvation state. To examine which effect is more dominant as the connectivity decreases, we directly compare the growth rate in ER and dBA networks with small connectivity $c$. Figure~\ref{fig:compare}A shows the growth rate under nutrient-poor conditions for $c=7$, $5$, and $3$. For $c=7$ and 5, the dBA network exhibits a larger growth rate than the ER network, indicating that the advantage of the power-law out-degree distribution persists. However, this advantage becomes less pronounced as $c$ decreases and almost disappears for $c=3$. This result suggests that, in sufficiently sparse networks, the effect of network sparsity becomes comparable to or even masks the advantage of the power-law out-degree distribution. The diminishing difference between the two network types can be understood from the Poisson distribution with small mean value $c$. As shown in Fig.~\ref{fig:compare}B, the probability of zero out-degree, $p_{v=0}^{\rm Poisson}$, increases rapidly as $c$ decreases and becomes appreciable around $c=5$. Because nodes with low out-degree stabilize the metabolic state under nutrient-poor conditions, their increasing abundance in sparse ER networks reduces the relative advantage conferred by the scale-free topology.}


\subsection*{Distributions in abundances of biomolecules}
\label{sec:distribution}

\begin{figure}[t]
\centering
\includegraphics[width=85mm]{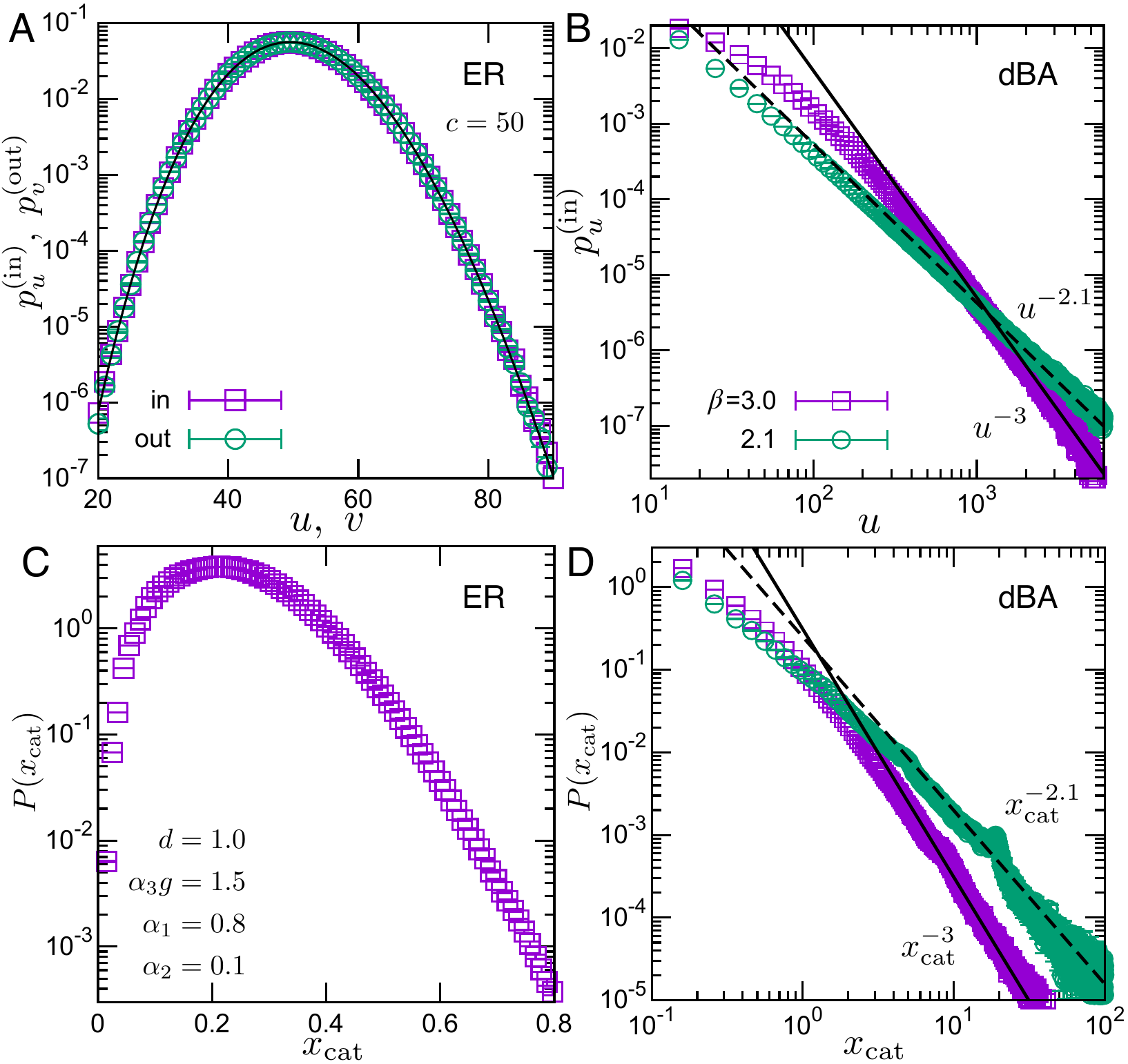}
\caption{
Semi-log plots of the distribution of (A) the in- and out-degree distributions, $\pin, \pout$, and (C) the abundance distribution of catalytic chemical species, $P(x_{\rm cat})$, for the ER network. The black curve in (A) represents the Poisson distribution. Log-log plots of (B) the in-degree distribution $\pin$ and (D) the abundance distribution of catalytic chemical species, $P(x_{\rm cat})$, for the dBA network with power-law in-degree distributions with the exponents $\beta = 3.0$ and $2.1$. Simulations are performed with parameters $N=5.0 \times 10^5$, $c=50$, $d=1.0$, $\alpha_3 g = 1.5$, $\alpha_1 = 0.8$ and $\alpha_2 = 0.1$ and averaged over 100 network realizations.
}
\label{fig:dist}
\end{figure}

We now turn to the distributions of chemical abundances. In particular, to understand the power-law behavior observed experimentally in biomolecules~\cite{furusawa2003zipf, ueda2004universality, lu2009investigation, sato2018power, lazzardi2023emergent}, we consider the abundance distributions of chemical species belonging to $\mathcal{G}_1$ and $\mathcal{G}_2$; that is, the abundance distributions of catalytic chemical species. 
\red{As shown above, the macroscopic quantities and phase behavior are independent of the in-degree distribution in dense networks. 
At the level of individual species, however, Eq.~\refbf{eq:fixed} indicates that the abundance of each chemical species depends on both its in-degree and out-degree. 
In particular, the power-law tail of the abundance distribution is determined primarily by the in-degree distribution rather than the out-degree distribution.}

One readily finds from Eq.~\refbf{eq:fixed} that the typical abundance of these species is proportional to their in-degree; hence, when the in-degree distribution follows a power law, the abundance distribution also follows a power law with a common exponent. To confirm this theoretical prediction, we numerically sample the abundance distributions of catalytic chemical species $P(x_{\rm cat})$ on the ER and dBA networks with in-degree power-law distributions. 

The degree distributions of ER and dBA networks are shown in Figs.~\ref{fig:dist}A and B. The in- and out-degree distributions of the ER network follow Poisson distributions. On the other hand, the in-degree distribution of the dBA network follows a power law $u^{-\beta}$, whose exponent can be controlled within the range $\beta>2$~\cite{barabasi2016network, dorogovtsev2000structure} (see {\it Materials and Methods} in detail). For the ER network, the abundance distribution does not have a fat tail, as shown in Fig.~\ref{fig:dist}C. This behavior reflects the in-degree Poisson distribution and is clearly distinguished from a power law. In contrast, for the dBA network, power-law tails are observed in the abundance distributions, with exponents identical to those of the in-degree distribution, as shown in Fig.~\ref{fig:dist}D. This agreement confirms the theoretical prediction that the abundance statistics of metabolic products directly reflect the underlying network topology.


\section*{Summary and Discussion}

In summary, we exactly solve a dense simple catalytic reaction network model with arbitrary degree distributions using DMFT. We show that the boundary of the metabolic-starvation transition sensitively depends on the out-degree distribution, whereas the boundary of the metabolic-overnutrition transition does not. In particular, power-law out-degree distributions, i.e., scale-free networks, suppress the transition to the starvation state. Although heterogeneous out-degree distribution is disadvantageous for cellular growth under nutrient-rich conditions, it becomes advantageous under nutrient-poor conditions, preventing the system from entering the starvation state. To test the validity of the DMFT results, we also performed numerical simulations on networks with finite connectivity, $c$. Under nutrient-rich conditions, the DMFT results agree well with simulations down to $c \approx 10$, whereas for $c < 10$ the metabolic state becomes unstable and transitions to the overnutrition state. In contrast, under nutrient-poor conditions, network sparsity further stabilizes the metabolic state. 

The mechanism underlying the disappearance of the metabolic-starvation transition on scale-free networks differs from those known in other systems, such as the Ising models~\cite{aleksiejuk2002ferromagnetic} and epidemic spreading models~\cite{pastor-satorras2001epidemic}. In those systems, the absence of a transition is typically attributed to highly connected hub nodes with large degrees. By contrast, in the present model, the transition disappears due to the presence of many nodes with extremely small out-degrees. In particular, unpenetrable metabolic products with small out-degrees relative to their in-degrees retain catalytic activity within the cell, preventing the system from losing mass and avoiding cellular shrinkage. This interpretation is consistent with the fact that the starvation transition is absent even for the uniform distribution with $r=1$, which does not possess a heavy tail.

Beyond the transition behavior, our theory also clarifies how network topology gives rise to biomolecular abundance distributions. We find that the abundance of metabolic products is proportional to the node's in-degree. Consequently, the abundance distribution exhibits a power-law tail when the in-degree distribution also follows a power law with the same exponent, which is confirmed by numerical simulations on dBA networks. Furusawa and Kaneko~\cite{furusawa2003zipf, furusawa2006evolutionary, furusawa2012adaptation} compared the chemical abundances in this model with gene expression levels measured by the amount of mRNAs. The gene expression levels are often presented as a rank-ordered distribution, with the exponent $\beta'$ typically between 0.9 and 1.0~\cite{furusawa2003zipf, ueda2004universality, lu2009investigation, lazzardi2023emergent}. The exponent $\beta$ of the original abundance distribution is related to $\beta'$ via the relation $\beta = (\beta'+1)/\beta'$. Using this relation yields $\beta \approx 2.1$, which is close to the experimentally observed exponent of approximately 2.2 for the degree distribution of metabolic networks~\cite{jeong2000the}.
\red{It should be noted, however, that the chemical species in this model represent coarse-grained biomolecular components. 
This comparison should therefore be interpreted as statistical trends, such as the presence of a power-law tail and its exponent, rather than as a one-to-one identification of model species with specific mRNAs or proteins.}

\red{We also note that the penetrable metabolic-products group $\mathcal{G}_2$ is not essential for the qualitative behavior discussed in this paper, including the transcritical bifurcation between the metabolic and overnutrition states, the suppression of the starvation transition by a power-law out-degree distribution, and the emergence of a power-law abundance distribution reflecting a power-law in-degree distribution. Indeed, even when $\alpha_2=0$, the metabolic-overnutrition transition remains a transcritical bifurcation. In this case, the phase boundary is reduced to $d_c=1/(\alpha_3 g-1)$, and the growth rate in the metabolic state is still largest for the Poisson out-degree distribution and smallest for the power-law out-degree distribution under the nutrient-rich condition (see SI Appendix, Fig.~S2A). The metabolic--starvation transition is also controlled by the out-degree distribution in the same manner as in the case with $\alpha_2>0$. For the Poisson out-degree distribution, the transition occurs at $\alpha_3 g=1-\alpha_1$, whereas for the uniform distribution with $r=1$ and for the power-law out-degree distribution, the growth rate exhibits an essential singularity and remains positive for any $\alpha_3 g>0$ (see SI Appendix, Fig.~S2B and S2C). Moreover, the mechanism for the power-law abundance distribution is unaffected by setting $\alpha_2=0$. The fixed-point abundance of chemical species in $\mathcal{G}_1$ remains proportional to the in-degree, and therefore a power-law in-degree distribution gives rise to a power-law abundance distribution with the same exponent.}

\red{Although the present study provides several insights into the relationship between network topology and metabolic dynamics, it has both methodological and modeling limitations. First, regarding the analytical methodology, our DMFT solution is exact in the thermodynamic and dense-connectivity limits for the random catalytic networks considered here. Real metabolic networks, however, contain a finite number of chemical species, typically on the order of $10^3$, and are sparsely connected, with only $2$--$3$ links per node~\cite{jeong2000the}. In this sparse regime, our simulations show noticeable deviations from the DMFT results. Moreover, the advantage conferred by a power-law out-degree distribution under nutrient-poor conditions becomes masked as $c$ decreases (see Fig.~\ref{fig:compare}A), which may be attributable to the increasing fraction of zero-out-degree nodes even when the out-degree distribution is Poisson (Fig.~\ref{fig:compare}B). This is consistent with the mechanism identified in the dense-limit theory, where low out-degree nodes help stabilize the metabolic state under nutrient-poor conditions. Nevertheless, this conclusion is still based on the present analysis of random catalytic-network model, and whether the same mechanism operates in real metabolic networks should be examined in future.}

\red{The second limitation concerns the abstract nature of the model itself. We represent all chemical reactions as catalytic conversions of the form $j+k\to i+k$, with uniform reaction rates. 
In addition, the cell-volume growth rate, $\mu$, is defined so as to maintain a constant total molecular concentration (see {\it Materials and Methods}), neglecting other relevant factors, such as the accumulation of inactive molecules and membrane elasticity. Whether our findings, including the suppression of the starvation transition and the relationship between the power-law distributions of in-degree and molecular abundance, persist in more realistic models of cellular growth remains an open question. A further limitation arises from our use of ensemble averages over typical realizations of networks with power-law degree distributions. Networks selected through evolution, however, may be typical, finely tuned structures~\cite{kaneko2022evolution, kikuchi2024phenotype}. For instance, hub chemical species and the species connected to them may not be selected at random~\cite{jeong2000the}. We also neglect network clustering, degree correlations between connected nodes, and correlations among the in-degree, out-degree, and catalytic degree of individual nodes, all of which may affect the dynamics. Even for the dBA networks, which exhibit weak degree correlations~\cite{fotouhi2013degree}, we observed a slight deviation from the DMFT results (Fig.~\ref{fig:starved}E). Despite these limitations, our exact analysis of a class of random metabolic-network models establishes a useful theoretical baseline against which dynamics on more realistic networks can be compared.}
 \\ \\ \noindent
 {\bf Data availability}\\
All data and code necessary to reproduce the results reported in this study are publicly available in the GitHub repository~(\href{https://github.com/kmitsumoto51/metabolic_scalefree}{https://github.com/kmitsumoto51/metabolic\_scalefree}).

 \begin{acknowledgments}
The authors would like to thank Kunihiko Kaneko, Chikara Furusawa, Yoshiyuki Kabashima, Hajime Yoshino, Toshinori Namba, Dominic K. Devlin, and Kei Oriyose for valuable discussions. This work was supported by JSPS KAKENHI Grant No. 25K17354, 24K00594, and No. 24H01931 from MEXT, Japan.
 \end{acknowledgments}

\appendix


\section*{Materials and Methods}
\subsection*{Dynamical equation and its nondimensionalization}
We present the detailed description of the dynamical equation, Eq.~\refbf{eq:dynamical1}, and the nondimensionalization of the variables and parameters.
Let $n_i$ be the number of molecules of chemical species $i$ in a cell of volume $V$, and let $\rho_i=n_i/V$ be its concentration. The time derivative of $\rho_i$ is written as
\begin{align}
\frac{d\rho_i}{dt} = \frac{1}{V}\frac{dn_i}{dt}-\rho_i \frac{1}{V}\frac{dV}{dt}.
\end{align}
The first term on the right-hand side represents changes in the molecule number due to catalytic reactions and exchange with the external environment. We write it as
\begin{align}
\frac{1}{V}\frac{dn_i}{dt} = \eta \sum_{j,k} C_{ij}^{k}\rho_j\rho_k-\eta \sum_{j,k} C_{ji}^{k}\rho_i\rho_k+d_i(g_i-\rho_i),
\label{eq:chem_reaction}
\end{align}
where $\eta$ is the reaction rate constant, $C_{ij}^{k}\in\{0,1\}$ is a binary tensor specifying whether the reaction $j+k\to i+k$ exists, $d_i$ is the exchange rate with the external environment, and $g_i$ is the equilibrium concentration outside the cell. We define the cell growth rate as $\mu=(1/V)dV/dt$. The concentration dynamics before nondimensionalization is therefore
\begin{align}
\frac{d\rho_i}{dt}=\eta \sum_{j,k} C_{ij}^{k}\rho_j\rho_k-\eta \sum_{j,k} C_{ji}^{k}\rho_i\rho_k+d_i(g_i-\rho_i)-\mu \rho_i.
\label{eq:dimensional_equation}
\end{align}
Let $N_{\rm tot}$ be the total number of molecules in the cell, i.e., $N_{\rm tot}=\sum_{i=1}^N n_i$. By taking the sum of Eq. \refbf{eq:chem_reaction}, one finds 
\begin{align}
\frac{dN_{\rm tot}}{dt}=V\sum_{i=1}^N d_i(g_i-\rho_i).
\end{align}
Now, we employ a minimal description of the cell-volume regulation that the total molecular concentration, $N_{\rm tot}/V$, remains constant in time. This assumption implies
\begin{align}
    \frac{1}{V}\frac{dV}{dt}=\frac{1}{N_{\rm tot}}\frac{dN_{\rm tot}}{dt}
\label{eq:relation_V_Ntot}
\end{align}
as follows from $\frac{d}{dt}\log(N_{\rm tot}/V)=0$. Using Eq.~\refbf{eq:relation_V_Ntot}, the growth rate is obtained as
\begin{align}
    \mu=\frac{V}{N_{\rm tot}}\sum_{i=1}^N d_i(g_i-\rho_i).
\end{align}
Thus, maintaining a constant total molecular concentration determines the growth rate as given above.

Let us introduce a dimensionless abundance $x_i = \rho_i/\rho_0$, where $\rho_0=N_{\rm tot}/VN$ is the concentration scale, to satisfy the normalization given by Eq.~\refbf{eq:constraint1}. We also nondimensionalize time as $\tilde{t} = c\eta\rho_0 t$, where $c$ is the mean connectivity of the reaction network. Substituting these rescaled variables into the dimensional concentration dynamics, we obtain the dimensionless dynamical equation as
\begin{align}
\frac{dx_i}{d\tilde{t}} = \frac{1}{c}\sum_{j,k}C_{ij}^k x_j x_k - \frac{1}{c}\sum_{j,k}C_{ji}^k x_i x_k + \tilde{d}_i(\tilde{g}_i-x_i) - \tilde{\mu}x_i,
\end{align}
where $\tilde{d}_i = d_i/c\eta\rho_0$, $\tilde {g}_i = g_i/\rho_0$, and $\tilde{\mu} = \mu/c\eta\rho_0 = \frac{1}{N}\sum_{i=1}^N \tilde{d}_i(\tilde {g}_i-x_i)$ are dimensionless permeability, equilibrium abundance outside the cell, cellular growth rate, respectively. Because we have assumed that $N_{\rm tot}/V$ is constant, $\tilde{d}_i$ and $\tilde{g}_i$ can be treated as constant parameters. In the main text, we denote them simply as $d_i$, $g_i$, and $\mu$, respectively.

\subsection*{Construction of sparse networks} 
We specify the catalytic reaction network by a binary adjacency tensor $C_{ij}^k \in \{0,1\}$. The tensor $C_{ij}^k$ is equal to 1 if the reaction $j+k\to i+k$ exists, and is equal to 0 otherwise. The catalytic reaction network is generated in three steps. We first construct a directed network described by an adjacency matrix $A_{ij} \in \{0,1\}$, where $A_{ij}=1$ means that the reaction channel $j\to i$ is allowed. Next, we assign to each node $i$ one of the following groups: unpenetrable metabolic product ($n=1$), penetrable metabolic product ($n=2$), or nutrient ($n=3$). The fractions of these species are fixed to be $\alpha_1$, $\alpha_2$, and $\alpha_3$, respectively, with $\alpha_1+\alpha_2+\alpha_3 = 1$. Finally, for each directed edge $A_{ij}=1$, we independently assign a catalyst species $k$ chosen uniformly at random from the set of non-nutrient species ($n=1,2$)~\cite{furusawa2003zipf}. In the case of Erd\H{o}s-R\'{e}nyi (ER) random networks, $A_{ij}$ is set to 1 with probability $c/N$ and to $0$ otherwise~\cite{erdos1959random, barabasi2016network, newman2010networks}. In the case of directed Barab\'{a}si-Albert (dBA) networks, $A_{ij}$ is generated via evolution with preferential attachment rule~\cite{barabasi2016network, dorogovtsev2000structure}. The preferential attachment rule to generate dBA network with a power-law in-degree distribution is as follows:
\begin{enumerate}
\item Prepare a complete directed network (without self-loops) consisting of $c+1$ nodes as the initial state. 
\item Add one new node to the network.
\item Create $c$ outgoing edges from the newly added node to existing nodes. Each endpoint node $i$ is chosen independently with probability $(u_i + A/c)/\sum_{i'} (u_{i'} + A/c)$, where $u_i$ denotes the current value of the in-degree of node $i$, and $A$ is the homogeneous attractiveness parameter. When selecting endpoints for the $c$ outgoing edges, we reject duplicates so that the new node connects to $c$ distinct existing nodes.
\item Repeat steps 2--3 until the total number of nodes reaches $N$.
\end{enumerate}
With the above preferential-attachment rule, the resulting stationary in-degree distribution has a power-law tail, $\pin \sim u^{-\beta}$ with $\beta = 2 + A/c$. Thus, the attractiveness parameter $A$ continuously tunes the tail exponent. Since each newly added node creates exactly $c$ outgoing edges, its out-degree distribution becomes $\pout = \delta_{v,c}$. If the roles of in-degree and out-degree is reversed, we can also generate dBA network with a power-law out-degree distribution.

\subsection*{Dynamics} 
For the simulation under high nutrient supply ($\alpha_3g>1$), we set $x_i = 1$ for all $i$ as the initial condition.
On the other hand, for the simulation under low nutrient supply ($\alpha_3g<1$), we set $x_i = (2-\alpha_3 g)/2\alpha_1$ for $i\in\mathcal{G}_1$ and $x_i = \alpha_3 g/2(\alpha_2 + \alpha_3)$ for $i\in\mathcal{G}_2,\mathcal{G}_3$ as the initial condition to avoid a negative growth rate at $t=0$. Then, the initial value of $\mu$ becomes $\alpha_3gd/2$. We solve Eq.~\refbf{eq:dynamical1} by using the fourth-order Runge-Kutta method. The time resolution $\Delta t = 0.02$ is used. 

\bibliography{fk_model_dmft}

\clearpage
\onecolumngrid
\setcounter{equation}{0}
\renewcommand{\theequation}{S\arabic{equation}}
\renewcommand{\thesection}{S\arabic{section}}
\begin{center}
{\large \textbf{Supporting information for ``Starvation suppression in dense scale-free metabolic networks: Dynamical mean-field analysis of catalytic reaction networks"}}
\end{center}
\vspace{5mm}
\section{Derivation for dynamical mean-field equation}

We consider a catalytic reaction network model consisting of $N$ chemical species~\cite{furusawa2003zipf, furusawa2006evolutionary}. The chemical species are divided into three groups: unpenetrable metabolic products, penetrable metabolic products, and nutrients, denoted by $\mathcal{G}_1, \mathcal{G}_2$ and $\mathcal{G}_3$, respectively. The fractions of chemical species belonging to these groups are denoted by $\alpha_1$, $\alpha_2$, and $\alpha_3$, respectively, satisfying $\alpha_1 + \alpha_2 + \alpha_3 = 1$. The model dynamics are described by the following differential equations:
\begin{align}
\dot{x}_i(t) = \frac{1}{c}\sum_{j =1}^N \sum_{k \notin \mathcal{G}_3} C_{ij}^k x_j(t) x_k(t) -  \frac{1}{c}\sum_{j =1}^N \sum_{k \notin \mathcal{G}_3}C_{ji}^k x_i(t) x_k(t) + d_i(g_i-x_i(t)) - \mu(t) x_i(t) + h_i(t) + \xi_i(t),
\label{eq:SM_model}
\end{align}
where the parameters $(d_i, g_i)$ are defined as $(0, 0)$ for $i \in \mathcal{G}_1$, $(d, 0)$ for $i \in \mathcal{G}_2$, and $(d, g)$ for $i \in \mathcal{G}_3$. $\mu = \frac{1}{N}\sum_{i =1}^N d_i(g_i-x_i)$ denotes the cell growth rate, and the term $- \mu(t) x_i(t)$ represents the dilution of chemical abundances due to the cell growth. The site-dependent field $h_i(t)$ is introduced for theoretical convenience and is set to zero at the end of the calculation. The variable $\xi_i(t)$ represents an uncorrelated Gaussian noise with zero mean and variance $\sigma^2$ (in the main text, we present results for $\xi_i(t) = 0$). The tensor $C_{ij}^k$ is equal to 1 if the reaction $j+k\to i+k$ exists, and is equal to 0 otherwise. We suppose that the nutrients cannot act as catalysts, i.e., $C_{ij}^k = 0$ for all $k\in \mathcal{G}_3$. $C_{ij}^k$ is generated using the configuration model~\cite{chung2002connected, newman2010networks}. Specifically, we first draw the in-degree $u_i$, out-degree $v_i$, and catalytic degree $w_i$ for all nodes $i$ independently from the distributions $\pin, \pout$ and $\pcat$. The probability distribution of $C_{ij}^k$ is then given by
\begin{align}
P(C_{ij}^k) = \frac{u_iv_jw_k}{(Nc)^2}\delta_{C_{ij}^k,1} + \qty(1-\frac{u_iv_jw_k}{(Nc)^2})\delta_{C_{ij}^k,0},
\label{eq:SM_cijk}
\end{align}
where $c$ is the mean connectivity, i.e., $\lim_{N\to\infty}\frac{1}{N}\sum_{i,j,k}C_{ij}^k = \sum_{u}\pin u = \sum_{v}\pout v = (\alpha_1+\alpha_2)\sum_{w}\pcat w = c$.

We solve the dynamics of Eq.~\refbf{eq:SM_model} on a network generated by  Eq.~\refbf{eq:SM_cijk} using dynamical mean-field theory~\cite{martin1973statistical, janssen1976on, dedominicis1978dynamics, sompolinsky1982relaxational, cugliandolo1993analytical, castellani2005spin, galla2024generating}. The theory is based on the generating functional
\begin{align}
\mathcal{Z}[\bm{h},\bm{\psi}] = \int \qty(\prod_{i=1}^N Dx_i) \mP{\bm{x}, \bm{h}}e^{i\int dt \sum_{i=1}^N x_i(t)\psi_i(t)}.
\label{eq:SM_GF_original}
\end{align}
The path probability distribution of $\bm{x}$ can be written as
\begin{align}
\mP{\bm{x}, \bm{h}} = &P_0(\bm{x}(0)) \int \qty(\prod_{i=1}^N D\xi_i)\mathcal{P}_{\sigma}[\bm{\xi}] \\
&\times\delta_{\rm F}\qty[\dot{x}_i(t)-\frac{1}{c}\sum_{j,k}C_{ij}^kx_j(t)x_k(t)-\frac{1}{c}\sum_{j',k'}C_{j'i}^kx_i(t)x_{k'}(t)-d_i(g_i-x_i(t))+\mu(t)x_i(t)-h_i(t)-\xi_i(t)],
\label{eq:SM_path_original}
\end{align}
where $\mathcal{P}_{\sigma}[\bm{\xi}]$ is the path probability distribution of the Gaussian noise
\begin{align}
\mathcal{P}_{\sigma}[\bm{\xi}] \sim \exp\qty(-\frac{1}{2\sigma^2}\int dt \sum_{i=1}^N\xi_i^2(t))
\label{eq:SM_path_noise}
\end{align}
and $P_0(\bm{x}(0))$ represents the joint initial distribution for all nodes $i$.
Using a functional integral representation of $\delta_{\rm F}$ and integrating out the Gaussian noise $\bm{\xi}_i$, we obtain
\begin{align}
\mathcal{Z}[\bm{h},\bm{\psi}] = \int \qty(\prod_{i=1}^N Dx_i D\hat{x}_i) P_0(\bm{x}(0))&\exp\qty(-\frac{i}{c}\sum_{i \neq j \neq k}C_{ij}^k \int dt \hat{x}_i(t) x_k(t) x_j(t) + \frac{i}{c}\sum_{i \neq j \neq k}C_{ji}^k \int dt \hat{x}_i(t) x_k(t) x_i(t)) \notag \\
\times &\exp\qty(i\int dt \sum_{i=1}^N A_i(x_i(t), \hat{x}_i(t), h_i(t), \psi_i(t))),
\label{eq:SM_GF1}
\end{align}
where a single-site action is defined as
\begin{align}
A_i(x_i(t), \hat{x}_i(t), h_i(t), \psi_i(t)) = x_i(t)\psi_i(t)+\frac{i \sigma^2}{2}\hat{x}_i^2(t) + \hat{x}_i(t)[\dot{x}_i(t)-d_i(g_i-x_i(t))+\mu(t)x_i(t)-h_i(t)].
\label{eq:SM_action}
\end{align}
From Eqs.~\refbf{eq:SM_GF1} and \refbf{eq:SM_action}, one finds that the fields $h_i(t)$ and $\psi_i(t)$ are coupled to the dynamical variables $\hat{x}_i(t)$ and $x_i(t)$, respectively. This leads to the following formulae,
\begin{align}
\expval{x_i(t)}_\xi &= (-i) \lim_{\bm{h}, \bm{\psi}\to 0} \frac{\delta Z[\bm{h}, \bm{\psi}]}{\delta \psi_i(t)}, \label{eq:SM_moment}  \\
\expval{\hat{x}_i(t)}_\xi &= i \lim_{\bm{h}, \bm{\psi}\to 0} \frac{\delta Z[\bm{h}, \bm{\psi}]}{\delta h_i(t)} = 0, \label{eq:SM_moment_hat}  \\
R(t_1,t_2) &= \frac{1}{N}\sum_{i=1}^N \lim_{\psi_i\to 0} \frac{\delta \expval{x_i(t_2)}_\xi}{\delta h_i(t_1)} = (-i) \lim_{\bm{h}, \bm{\psi}\to 0} \frac{1}{N}\sum_{i=1}^N\frac{\delta^n Z[\bm{h}, \bm{\psi}]}{\delta h_i(t_1) \delta \psi_i(t_2)} = \frac{-i}{N}\sum_{i=1}^N \expval{\hat{x}_i(t_1)x_i(t_2)}_\xi, \label{eq:SM_response} 
\end{align}
where $\expval{\cdots}_\xi$ denotes the average over the noise and $R(t_1,t_2)$ represents the response function. Because of the causality, $R(t_1,t_2) = 0$ if $t_2 \le t_1$, although $\lim_{t_2 \to t_1^-}R(t_1,t_2) \neq 0$.
We also note that the generating functional satisfies the normalization condition $Z[\bm{h}, 0] = 1$, leading to $\expval{\hat{x}_i(t)}_\xi=0$, as given in Eq.~\refbf{eq:SM_moment_hat}.

Let us evaluate the generating functional $\ovl{\mathcal{Z}[\bm{h},\bm{\psi}]}$ averaged over the network topology $\qty{C_{ij}^k}$. In general, this averaged generating functional can be written as
\begin{align}
\ovl{\mathcal{Z}[\bm{h},\bm{\psi}]} = \int \qty(\prod_{i=1}^N Dx_i D\hat{x}_i) P_0(\bm{x}(0)) \Delta [\bm{x}, \hat{\bm{x}}] \exp\qty(i\int dt \sum_{i=1}^N A_i(x_i(t), \hat{x}_i(t), h_i(t), \psi_i(t))),
\label{eq:SM_GF2}
\end{align}
where
\begin{align}
\Delta [\bm{x}, \hat{\bm{x}}] &= \ovl{ \exp\qty(-\frac{i}{c}\sum_{i \neq j \neq k}C_{ij}^k \int dt \hat{x}_i(t) x_k(t) x_j(t) + \frac{i}{c}\sum_{i \neq j \neq k}C_{ji}^k \int dt \hat{x}_i(t) x_k(t) x_i(t))} \notag \\
&= \ovl{ \prod_{i \neq j \neq k} \exp\qty(-\frac{i}{c}C_{ij}^k \int dt  ( \hat{x}_i(t) -  \hat{x}_j(t))x_k(t) x_j(t))} \nonumber \\
&= \exp \qty[\sum_{i \neq j \neq k} \ln \qty(1+\frac{u_iv_jw_k}{(Nc)^2}\qty(\exp \qty( -\frac{i}{c}\int dt ( \hat{x}_i(t) -  \hat{x}_j(t))x_k(t) x_j(t))-1))]
\label{eq:SM_delta1}
\end{align}
To decouple the interaction terms between different sites in $\Delta [\bm{x}, \hat{\bm{x}}]$, we introduce the following functional order parameters~\cite{mimura2009parallel, poley2025interaction, metz2025dynamical}: 
\begin{align}
U_{nu}[\hat{x}] &= \frac{1}{\alpha_n \pin N}\sum_{i \in \mathcal{G}_n} \delta_{u_i,u} \delta_{\rm F}[\hat{x}(t)-\hat{x}_i(t)]~~~(n=1,2,3), \label{eq:SM_order1} \\
V_{nv}[x, \hat{x}] &= \frac{1}{\alpha_n \pout N}\sum_{i \in \mathcal{G}_n} \delta_{v_i,v} \delta_{\rm F}[x(t)-x_i(t)]\delta_{\rm F}[\hat{x}(t)-\hat{x}_i(t)]~~~(n=1,2,3), \label{eq:SM_order2}\\
W_{nw}[x] &= \frac{1}{\alpha_n \pcat N}\sum_{i \in \mathcal{G}_n} \delta_{w_i,w} \delta_{\rm F}[x(t)-x_i(t)]~~~(n=1,2).\label{eq:SM_order3}
\end{align}
For notational convenience, we also introduce $W_{3w}[x] = 0$.
Using these order parameters, we can write $\Delta [\bm{x}, \hat{\bm{x}}]$ as
\begin{align}
\Delta [\bm{x}, \hat{\bm{x}}] = \exp \qty[N\sum_{u,v,w}\pin \pout \pcat\int D\hat{x} Dy D\hat{y} Dz (\sum_{n=1}^3 \alpha_n U_{nu}[\hat{x}])(\sum_{n=1}^3 \alpha_n V_{nv}[y, \hat{y}])(\sum_{n=1}^3 \alpha_n W_{nw}[z])f_{uv}^w[\hat{x}, y, \hat{y}, z]],
\label{eq:SM_delta2}
\end{align}
where
\begin{align}
f_{uv}^w[\hat{x}, y, \hat{y}, z] = N^2 \ln \qty(1+\frac{uvw}{(Nc)^2}\qty(\exp \qty(-\frac{i}{c}\int dt (\hat{x}(t) -  \hat{y}(t))y(t) z(t))-1)).
\label{eq:SM_fuvw1}
\end{align}
In the dense connectivity limit, the leading terms in powers of $1/c$ are relevant. Hence, by expanding the exponential in Eq.~\refbf{eq:SM_fuvw1} with respect to $1/c$ and taking the thermodynamic limit $N\to\infty$ and, we obtain
\begin{align}
f_{uv}^w[\hat{x}, y, \hat{y}, z] = -i\frac{uvw}{c^3}\int dt (\hat{x}(t) -  \hat{y}(t))y(t) z(t).
\label{eq:SM_fuvw2}
\end{align}
By inserting the following identities,
\begin{align}
1 &= \int DU_{nu} \delta_{\rm F}\Bigl[\alpha_n \pin NU_{nu}[\hat{x}]-\sum_{i \in \mathcal{G}_n}\delta_{u_i,u} \delta_{\rm F}[\hat{x}(t)-\hat{x}_i(t)]\Bigr] \nonumber \\
&= \frac{1}{2\pi} \int DU_{nu} D\hat{U}_{nu} e^{i\alpha_n \pin N\int D\hat{x} \hat{U}_{nu}[\hat{x}] U_{nu}[\hat{x}]-i\sum_{i \in \mathcal{G}_n} \delta_{u_i,u}\hat{U}_{nu}[\hat{x}_i]}, \label{eq:SM_identity1} \\
1 &= \int DV_{nv} \delta_{\rm F}\Bigl[\alpha_n \pout NV_{nv}[x,\hat{x}]-\sum_{i \in \mathcal{G}_n}\delta_{v_i,v} \delta_{\rm F}[x(t)-x_i(t)]\delta_{\rm F}[\hat{x}(t)-\hat{x}_i(t)]\Bigr] \nonumber \\
&= \frac{1}{2\pi} \int DV_{nv} D\hat{V}_{nv} e^{i \alpha_n \pout N\int Dx D\hat{x} \hat{V}_{nv}[x, \hat{x}] V_{nv}[x, \hat{x}]-i\sum_{i \in \mathcal{G}_n}\delta_{u_i,u}\hat{V}_{nv}[x_i, \hat{x}_i]}, \label{eq:SM_identity2} \\
1 &= \int DW_{nw} \delta_{\rm F}\Bigl[\alpha_n \pcat NW_{nw}[x]-\sum_{i \in \mathcal{G}_n}\delta_{u_i,u} \delta_{\rm F}[x(t)-x_i(t)]\Bigr] \nonumber \\
&= \frac{1}{2\pi}\int DW_{nw} D\hat{W}_{nw} e^{i \alpha_n \pcat N\int Dx \hat{W}_{nw}[x] W_{nw}[x]-i\sum_{i \in \mathcal{G}_n}\delta_{w_i,w} \hat{W}_{nw}[x_i]},
\end{align}
into the averaged generating functional $\ovl{Z[\bm{h},\bm{\psi}]}$, We obtain
\begin{align}
\ovl{\mathcal{Z}[\bm{h},\bm{\psi}]} = \int \qty(\prod_{n,u}DU_{nu}D\hat{U}_{nu})\qty(\prod_{n,v}DV_{nv}D\hat{V}_{nv})\qty(\prod_{n,w}DW_{nw}D\hat{W}_{nw})\exp(N\Phi[\bm{U},\hat{\bm{U}},\bm{V}, \hat{\bm{V}}, \bm{W}, \hat{\bm{W}}]),
\label{eq:SM_GF3}
\end{align}
where
\begin{align}
\Phi &= \sum_{u,v,w}\pin \pout \pcat\int D\hat{x} Dy D\hat{y} Dz (\sum_{n=1}^3 \alpha_n U_{nu}[\hat{x}])(\sum_{n=1}^3 \alpha_n V_{nv}[y, \hat{y}])(\sum_{n=1}^3 \alpha_n W_{nw}[z])f_{uv}^w[\hat{x}, y, \hat{y}, z] \nonumber \\
&+ i\sum_{n=1}^3\alpha_n \qty(\sum_u \pin \int D\hat{x} U_{nu}[\hat{x}]\hat{U}_{nu}[\hat{x}] + \sum_v \pout \int Dx D\hat{x} V_{nv}[x, \hat{x}]\hat{V}_{nv}[x, \hat{x}] + \sum_w \pcat \int Dx W_{nw}[x]\hat{W}_{nw}[x]) \nonumber \\
&+\sum_{n,u,v,w}\alpha_n \pin \pout \pcat \ln \Biggl(\int Dx D\hat{x} P_0(x(0)) \nonumber \\ 
&~~~~~~~~~~~~~~~~~~~~~~~~~~\times\exp \qty[i\int dt A_n(x(t), \hat{x}(t), h_{nuv}^w(t), \psi_{nuv}^w(t)) - i\hat{U}_{nu} [\hat{x}] - i\hat{V}_{nv} [x, \hat{x}] - i\hat{W}_{nw} [x]]\Biggr).
\label{eq:SM_Phi}
\end{align}
Here, we assume that $h_i(t)$ and $\psi_i(t)$ depend only on the in-degree $u$, out-degree $v$, catalytic degree $w$ and group $\mathcal{G}_n$ of the $i$-th chemical species; this assumption leads to the factorized form of the last term in Eq.~\refbf{eq:SM_Phi}.

The integral in Eq.~\refbf{eq:SM_GF3} can be evaluated by applying the saddle-point method in the large $N$ limit, and then the resulting saddle-point equations read
\begin{align}
\hat{U}_{nu}[\hat{x}] &= i \sum_{v,w}\pout \pcat \int Dy D\hat{y}Dz (\sum_{n=1}^3 \alpha_n V_{nv}[y, \hat{y}])(\sum_{n=1}^3 \alpha_n W_{nw}[z])f_{uv}^w[\hat{x}, y, \hat{y}, z] \equiv \hat{U}_{u}[\hat{x}], \label{eq:SM_saddle1} \\
\hat{V}_{nv}[y,\hat{y}] &= i \sum_{u,w}\pin \pcat \int D\hat{x} Dz (\sum_{n=1}^3 \alpha_n U_{nu}[\hat{x}]) (\sum_{n=1}^3 \alpha_n W_{nw}[z])f_{uv}^w[\hat{x}, y, \hat{y}, z] \equiv \hat{V}_{v}[y,\hat{y}], \label{eq:SM_saddle2} \\
\hat{W}_{nw}[z] &= i \sum_{u,v} \pin \pout \int D\hat{x} Dy D\hat{y} (\sum_{n=1}^3 \alpha_n U_{nu}[\hat{x}]) (\sum_{n=1}^3 \alpha_n V_{nv}[y, \hat{y}]) f_{uv}^w[\hat{x}, y, \hat{y}, z] \equiv \hat{W}_{w}[z], \label{eq:SM_saddle3} \\
U_{nu}[\hat{x}] &= \sum_{v,w} \pout \pcat \int Dx \gamma[x,\hat{x}|n,u,v,w], \label{eq:SM_saddle4} \\
V_{nv}[y, \hat{y}] &= \sum_{u,w} \pout \pcat \gamma[y,\hat{y}|n,u,v,w], \label{eq:SM_saddle5} \\
W_{nw}[z] &= \sum_{u,v} \pin \pout \int D\hat{z} \gamma[z,\hat{z}|n,u,v,w], \label{eq:SM_saddle6}
\end{align}
where
\begin{align}
\gamma[x,\hat{x}|n,u,v,w] = \frac{\exp \qty[i\int dt A_n(x(t), \hat{x}(t), h_{nuv}^w(t), \psi_{nuv}^w(t)) - i\hat{U}_{u}[\hat{x}] - i\hat{V}_{v}[x,\hat{x}] - i\hat{W}_{w}[x]]}{\int Dx D\hat{x} \exp \qty[i\int dt A_n(x(t), \hat{x}(t), h_{nuv}^w(t), \psi_{nuv}^w(t)) - i\hat{U}_{u}[\hat{x}] - i\hat{V}_{v}[x,\hat{x}] - i\hat{W}_{w}[x]]}.
\label{eq:SM_gamma}
\end{align}
It should be noted that the saddle-point equations for the conjugate (hatted) order parameters do not depend on the species group $n$; therefore, we introduce the shorthand notation, $\hat{U}_{u}[\hat{x}], \hat{V}_{v}[y, \hat{y}]$, and $\hat{W}_{w}[z]$. 

To interpret the probability distribution $\gamma[x,\hat{x}|n,u,v,w]$, we define the expectation with respect to $\gamma[x,\hat{x}|n,u,v,w]$ as
\begin{align}
\expval{\cdots}_{nuv}^w = \int Dx D\hat{x} \cdots \gamma[x,\hat{x}|n,u,v,w].
\end{align}
By calculating the functional derivatives of the expression for the generating functional $\ovl{\mathcal{Z}[\bm{h},\bm{\psi}]}$ in Eq.~\refbf{eq:SM_GF3} with respect to $\psi_{nuv}^w(t)$ and $h_{nuv}^w(t)$, we find the following relations:
\begin{align}
\expval{x(t)}_{nuv}^w = \int Dx D\hat{x} x(t) \gamma[x,\hat{x}|n,u,v,w] 
&=  (-i) \lim_{\bm{h},\bm{\psi}\to 0} \frac{\delta \ovl{\mathcal{Z}[\bm{h},\bm{\psi}]}}{\delta \psi_{nuv}^w(t)} \nonumber \\
&=  (-i) \lim_{\bm{h},\bm{\psi}\to 0} \frac{1}{\alpha_n \pin \pout \pcat N}\sum_{i \in \mathcal{G}_n} \delta_{u_i,u} \delta_{v_i,v} \delta_{w_i,w} \frac{\delta \ovl{\mathcal{Z}[\bm{h},\bm{\psi}]}}{\delta \psi_i(t)} \nonumber \\
&= \frac{1}{\alpha_n \pin \pout \pcat N}\sum_{i \in \mathcal{G}_n} \delta_{u_i,u} \delta_{v_i,v} \delta_{w_i,w} \expval{x_i(t)}_\xi, \label{eq:SM_moment2} \\
\expval{\hat{x}(t)}_{nuv}^w = \int Dx D\hat{x} \hat{x}(t) \gamma[x,\hat{x}|n,u,v,w] 
&=  i \lim_{\bm{h},\bm{\psi}\to 0} \frac{\delta \ovl{\mathcal{Z}[\bm{h},\bm{\psi}]}}{\delta h_{nuv}^w(t)} \nonumber \\
&=  i \lim_{\bm{h},\bm{\psi}\to 0} \frac{1}{\alpha_n \pin \pout \pcat N}\sum_{i \in \mathcal{G}_n} \delta_{u_i,u} \delta_{v_i,v} \delta_{w_i,w} \frac{\delta \ovl{\mathcal{Z}[\bm{h},\bm{\psi}]}}{\delta h_i(t)} \nonumber \\
&= \frac{1}{\alpha_n \pin \pout \pcat N}\sum_{i \in \mathcal{G}_n} \delta_{u_i,u} \delta_{v_i,v} \delta_{w_i,w} \expval{\hat{x}_i(t)}_\xi = 0, \label{eq:SM_moment_hat2} \\
\expval{x(t)\hat{x}(t')}_{nuv}^w = \int Dx D\hat{x} x(t) \hat{x}(t') \gamma[x,\hat{x}|n,u,v,w] &= \frac{1}{\alpha_n \pin \pout \pcat N}\sum_{i \in \mathcal{G}_n} \delta_{u_i,u} \delta_{v_i,v} \delta_{w_i,w} \expval{x(t)\hat{x}_i(t')}_\xi. \label{eq:SM_responce2}
\end{align}
Comparing these relations with Eqs.~\refbf{eq:SM_moment}-\refbf{eq:SM_response}, one can regard $\int D\hat{x} \gamma[x,\hat{x}|n,u,v,w]$ as the path probability describing the typical dynamics of the abundance of the chemical species belonging to $\mathcal{G}_n$, conditioned on the in-degree $u$, out-degree $v$, and catalytic degree $w$.
Using  Eqs.~\refbf{eq:SM_moment2}-\refbf{eq:SM_responce2} and substituting Eqs.~\refbf{eq:SM_fuvw2}, \refbf{eq:SM_saddle4}-\refbf{eq:SM_saddle6} into Eq.~\refbf{eq:SM_saddle3}, we find
\begin{align}
\hat{W}_w[z] &= \sum_{u,v}\pin \pout \frac{uvw}{c^3} \Biggl[\int dt (\sum_{n=1}^3\alpha_n \sum_{v', w'} p_{v'}^{\rm (out)} p_{w'}^{\rm (cat)} \expval{\hat{x}(t)}_{nuv'}^{w'})(\sum_{n=1}^3\alpha_n \sum_{v', w'} p_{u'}^{\rm (in)} p_{w'}^{\rm (cat)} \expval{y(t)}_{nu'v}^{w'})z(t) \nonumber \\
&~~~~~~~~~~~~~~~~~~~~~~~~~~~~~~~~~~~~~~~~~~~~~~~~~~~~~~~~~~~~~~~~~~~~~- (\sum_{n=1}^3\alpha_n \sum_{u', w'} p_{u'}^{\rm (in)} p_{w'}^{\rm (cat)}\expval{y(t)\hat{y}(t)}_{nu'v}^{w'})z(t)\Biggr] \nonumber \\
&= 0.
\label{eq:SM_w_hat}
\end{align}
Here, we use $\expval{\hat{x}(t)}_{nuv}^w=0$ and $\expval{\hat{x}(t)x(t)}_{nuv}^w=0$.
Vanishing of $\hat{W}_w[z]$ implies that $\int D\hat{x} \gamma[x,\hat{x}|n,u,v,w]$ does not depend on the catalytic degree $w$, i.e., the effective dynamics of the chemical abundance is independent of $w$. This result has a clear physical interpretation. The catalyzed chemical species do not provide feedback to the catalytic chemical species through the catalytic activity. Thus, the number of catalytic links $w$ does not affect the effective dynamics of the chemical abundance and then the probability measure and averages can be simplified as $\gamma[x,\hat{x}|n,u,v,w] \to \gamma[x,\hat{x}|n,u,v]$ and $\expval{\cdots}_{nuv}^w \to \expval{\cdots}_{nuv}$.

Similarly to $\hat{W}_w[z]$, the other hatted order parameters can be calculated as
\begin{align}
\hat{U}_u[\hat{x}] &= \sum_{v,w}\pout \pcat \frac{uvw}{c^3}\Biggl[\int dt \hat{x}(t)(\sum_{n=1}^3\alpha_n\sum_{u'} p_{u'}^{\rm (in)}\expval{y(t)}_{nu'v})(\sum_{n=1}^2\alpha_n\sum_{u', v'} p_{u'}^{\rm (in)} p_{v'}^{\rm (out)}\expval{z(t)}_{nu'v'}) \nonumber \\
&~~~~~~~~~~~~~~~~~~~~~~~~~~~~~~~~~~~~~~~~~~-(\sum_{n=1}^3\alpha_n\sum_{u'} p_{u'}^{\rm (in)}\expval{\hat{y}(t)y(t)}_{nu'v})(\sum_{n=1}^2\alpha_n\sum_{u', v'} p_{u'}^{\rm (in)} p_{v'}^{\rm (out)}\expval{z(t)}_{nu'v'}) \Biggr] \nonumber \\
&= \frac{u}{c} \int dt \hat{x}(t) m_{\rm in}(t) \mcat(t), \label{eq:SM_u_hat} \\
\hat{V}_v[y, \hat{y}] &= \sum_{u,w} \pin \pcat \frac{uvw}{c^3}\Biggl[\int dt (\sum_{n=1}^3\alpha_n\sum_{v'} p_{v'}^{\rm (out)}\expval{\hat{x}(t)}_{nuv'})y(t)(\sum_{n=1}^2\alpha_n\sum_{u',v'} p_{u'}^{\rm (in)} p_{v'}^{\rm (out)}\expval{z(t)}_{nu'v'})\nonumber \\
&~~~~~~~~~~~~~~~~~~~~~~~~~~~~~~~~~~~~~~~~~~~~~~~~~~~~~~~~~~~~~~~~~~~~~~~-\hat{y}(t)y(t)(\sum_{n=1}^2\alpha_n\sum_{u',v'} p_{u'}^{\rm (in)} p_{v'}^{\rm (out)}\expval{z(t)}_{nu'v'})\Biggr] \nonumber \\
&=- \frac{v}{c} \int dt \hat{y}(t) y(t) \mcat(t), \label{eq:SM_v_hat}  
\end{align}
where
\begin{align}
m_{\rm in}(t) &= \sum_{n, u, v} \frac{v}{c} \alpha_n \pin \pout \expval{x(t)}_{nuv}, \label{eq:SM_min} \\
\mcat(t) &= \frac{\sum_{u,v} \pin \pout \qty(\alpha_1 \expval{x(t)}_{1uv} + \alpha_2 \expval{x(t)}_{2uv})}{\alpha_1 + \alpha_2}. \label{eq:SM_mcat}
\end{align}
Substituting Eqs.~\refbf{eq:SM_u_hat} and \refbf{eq:SM_v_hat} into $\gamma[x,\hat{x}|n,u,v]$, we obtain
\begin{align}
\gamma[x,\hat{x}|n,u,v] &= \frac{\exp \qty[i\int dt A_n(x(t), \hat{x}(t), h_{nuv}(t), \psi_{nuv}(t))-i\hat{x}(t)(\frac{u}{c}m_{\rm in}(t) \mcat(t)-\frac{v}{c}x(t)\mcat(t))]}{\int DxD\hat{x}\exp \qty[i\int dt A_n(x(t), \hat{x}(t), h_{nuv}(t), \psi_{nuv}(t))-i\hat{x}(t)(\frac{u}{c}m_{\rm in}(t) \mcat(t)-\frac{v}{c}x(t)\mcat(t))]}.
\end{align}
The path probability of the effective dynamical equation for the abundance of the chemical species belonging to $\mathcal{G}_n$, conditioned on the in-degree $u$ and out-degree $v$, can then be written as
\begin{align}
\nonumber
\mP{x_{nuv}, h_{nuv}} = &P_0(x_{nuv}(0))\int D\xi_{nuv}\mathcal{P}_\sigma [\xi_{nuv}] \\
&\times\delta_{\rm F}\qty[\dot{x}_{nuv}(t)-\frac{u}{c} m_{\rm in}(t) \mcat (t) + \frac{v}{c} \mcat (t) x_{nuv}(t)- d_n (g_n - x_{nuv}(t))+\mu(t)x_{nuv}(t) -h_{nuv}(t) - \xi_{nuv}(t)].
\end{align}
Hence, we finally obtain the effective dynamical equations as
\begin{align}
\dot{x}_{nuv}(t) = \frac{u}{c} m_{\rm in}(t) \mcat (t) - \frac{v}{c} \mcat (t) x_{nuv}(t) + d_n (g_n - x_{nuv}(t)) - \mu(t)x_{nuv}(t) + h_{nuv}(t) + \xi_{nuv}(t).
\label{eq:SM_dmfteq}
\end{align}
In the main text, we analyze the case of $h_{nuv}(t) = \xi_{nuv}(t) = 0$.


\section{Linear stability analysis}
\label{sec:stability}
To examine the stability of the non-metabolic solution,  we linearize the DMFT equation, given by Eq.~\refbf{eq:SM_dmfteq}, around the fixed point by introducing a small perturbation $\xi_{nuv}(t)$~\cite{opper1992phase, roy2019numerical, galla2024generating}. The perturbation is assumed to be an independent white Gaussian noise for each $n$, $u$, and $v$. We define the deviations from the fixed point as $\delta x_{nuv}(t) = x_{nuv}(t) - x_{nuv}^*$, $\delta \mcat(t) = \mcat(t) - \mcat^*$, $\delta m_{\rm in}(t) = m_{\rm in}(t) - m_{\rm in}^*$, and $\delta \mu(t) = \mu(t) - \mu^*$. If the solution is stable, all these deviations fluctuate around zero in the long-time limit.

The linearized dynamics around a given fixed point are written as
\begin{align}
\delta \dot{x}_{nuv}(t) = &\frac{u}{c}(m_{\rm in}^*\delta \mcat(t) + \delta m_{\rm in}(t) \mcat^*) - \frac{v}{c}(\mcat^*\delta x_{nuv}(t) + \delta \mcat(t) x_{nuv}^*) -(d_n+\mu^*)\delta x_{nuv}(t)-\delta\mu(t) x_{nuv}^* + \xi_{nuv}.
\label{eq:linear}
\end{align}
Denoting by $\tilde{f}(\omega)$ the Fourier transform of a function $f(t)$, Eq.~\refbf{eq:linear} can be rewritten as
\begin{align}
\delta \tilde{x}_{nuv}(\omega) = &\frac{\frac{u}{c}(m_{\rm in}^*\delta \tilde{m}_{\rm cat}(\omega) + \delta \tilde{m}_{\rm in}(\omega) \mcat^*)}{i \omega + d_n + \mu^* + \frac{v}{c}\mcat^*} -\frac{(\frac{v}{c}\delta \tilde{m}_{\rm cat}(\omega) - \delta \tilde{\mu}(\omega))x_{nuv}^* + \tilde{\xi}_{nuv}(\omega)}{i \omega + d_n + \mu^* + \frac{v}{c}\mcat^*}.
\label{eq:linear_fourier}
\end{align}
Substituting the non-metabolic solution into Eq.~\refbf{eq:linear_fourier}, we obtain the following expressions for
\begin{align}
\delta \tilde{x}_{1uv}(\omega) &= \frac{\frac{u}{c}\delta \tilde{m}_{\rm cat}(\omega) + \tilde{\xi}_{1uv}(\omega)}{i \omega + d(\alpha_3 g -1)}, \\
\delta \tilde{x}_{2uv}(\omega) &= \frac{\frac{u}{c}\delta \tilde{m}_{\rm cat}(\omega) + \tilde{\xi}_{2uv}(\omega)}{i \omega + d\alpha_3 g}, \\
\delta \tilde{x}_{3uv}(\omega) &= \frac{\frac{u-v/\alpha_3}{c}\delta \tilde{m}_{\rm cat}(\omega) - \delta \tilde{\mu}(\omega)/\alpha_3 + \tilde{\xi}_{3uv}(\omega)}{i \omega + d\alpha_3 g}.
\end{align}
We then square these expressions and average over the Gaussian noise $\xi_{nuv}$ to obtain
\begin{align}
\expval{|\delta \tilde{x}_{1uv}(\omega)|^2} &= \frac{(\frac{u}{c})^2 \expval{|\delta \tilde{m}_{\rm cat}(\omega)|^2} + I_{1uv}}{\omega^2 + d^2(\alpha_3 g -1)^2}, \label{eq:delta_x1uv} \\
\expval{|\delta \tilde{x}_{2uv}(\omega)|^2} &= \frac{(\frac{u}{c})^2 \expval{|\delta \tilde{m}_{\rm cat}(\omega)|^2} + I_{2uv}}{\omega^2 + d^2\alpha_3^2 g^2}, \label{eq:delta_x2uv} \\
\expval{|\delta \tilde{x}_{3uv}(\omega)|^2} &= \frac{\expval{|\frac{(u-v/\alpha_3)\delta \tilde{m}_{\rm cat}}{c} - \frac{\delta \tilde{\mu}(\omega)}{\alpha_3}|^2} + I_{3uv}}{\omega^2 + d^2\alpha_3^2 g^2}, \label{eq:delta_x3uv}
\end{align}
where $\expval{\cdots}$ denotes the average over $\xi_{nuv}$ and $I_{nuv} = \expval{|\tilde{\xi}_{nuv}|^2}$. In the limit $\omega \to 0$, one finds that $\expval{|\delta \tilde{x}_{1uv}(\omega)|^2}$ diverges at $\alpha_3 g = 1$, whereas $\expval{|\delta \tilde{x}_{2uv}(\omega)|^2}$ and $\expval{|\delta \tilde{x}_{3uv}(\omega)|^2}$ diverge at $\alpha_3 g = 0$. Hence, for any permeability $d$, the non-metabolic state becomes unstable at $\alpha_3 g = 1$ if $\alpha_1 > 0$, and otherwise at $\alpha_3 g = 0$.
 
The non-metabolic state can be unstable even for $\alpha_3 g > 1$ if the numerators in Eqs.~\refbf{eq:delta_x1uv}-\refbf{eq:delta_x3uv} diverge. From the definition of the average catalytic abundance $\mcat$, we obtain the following self-consistent relation for
\begin{align}
\delta \tilde{m}_{\rm cat}(\omega) = &\frac{1}{\alpha_1 + \alpha_2}\Big(\alpha_1 \frac{\delta \tilde{m}_{\rm cat}(\omega) + \tilde{\xi}_{1}(\omega)}{i \omega + d(\alpha_3 g -1)} + \alpha_2 \frac{\delta \tilde{m}_{\rm cat}(\omega) + \tilde{\xi}_{2}(\omega)}{i \omega + d\alpha_3 g} \Big), \label{eq:delta_mcat}
\end{align}
where $\tilde{\xi}_{n} = \sum_{u,v} \pin \pout \tilde{\xi}_{nuv}(\omega)$. Evaluating the zero-frequency limit of $\expval{|\delta \tilde{m}_{\rm cat}(\omega)|^2}$, we find
\begin{align}
\expval{|\delta \tilde{m}_{\rm cat}(0)|^2} \propto \frac{1}{d^2[d(\alpha_3 g -1)\alpha_3 g - \frac{\alpha_1 \alpha_3 g + \alpha_2(\alpha_3 g -1)}{\alpha_1 + \alpha_2}]^2}.
\end{align}
The divergence of this quantity determines the critical permeability, $d_{\rm c}$, given in Eq. {\bf 9} in the main text. Hence, for $d < d_{\rm c}$, the fluctuations $\expval{|\delta \tilde{x}_{nuv}(\omega)|^2}$ diverge for all $n$, $u$, and $v$, leading to the instability of the non-metabolic solution, as shown in Fig.~2A in the main text.


\section{Metabolic solution for the Poisson degree distribution}
\label{sec:metabo_poisson}
In the dense connectivity limit, the Poisson out-degree distribution reduces to the Kronecker delta, $\pout = \delta_{c,v}$.
As a result, $X(a, b) = \sum_v \pout/(a + \frac{v}{c}b) = 1/(a+b)$ and $m_{\rm in}^* = 1$. Substituting these relations into Eq.~{\bf 7} in the main text, we obtain
\begin{align}
\alpha_1+ \alpha_2 = \frac{\alpha_1}{\mu^*+\mcat^*} + \frac{\alpha_2}{\mu^*+\mcat^* + d}.
\end{align}
Solving this equation yields
\begin{align}
\mu^* + \mcat^* &= \frac{1-d + \sqrt{(1-d)^2+4\alpha_1 d/(\alpha_1+\alpha_2)}}{2} \nonumber \\
& \equiv \phi(d, \alpha_1,\alpha_2).
\label{eq:Phi}
\end{align}
Here, we have used the fact that $\phi(d, \alpha_1,\alpha_2)$ must be positive.
Substituting Eq.~\refbf{eq:Phi} into Eqs.~{\bf 7} and {\bf 8} in the main text, we finally obtain the growth rate and the average catalytic abundance as
\begin{align}
\mu^* &= \frac{d(\alpha_3 g + \alpha_1 -1)\phi(d, \alpha_1,\alpha_2)}{\phi(d, \alpha_1,\alpha_2) + \alpha_1 d}, \\
\mcat^* &= \frac{(\phi(d, \alpha_1,\alpha_2)-d(\alpha_3 g-1))\phi(d, \alpha_1,\alpha_2)}{\phi(d, \alpha_1,\alpha_2) + \alpha_1 d},
\end{align}
which are shown in Figs.~3A and B in the main text, respectively. Finally, substituting the critical point $d_{\rm c}$, given by Eq.~{\bf 10} in the main text, into the above expressions, one can verify that the metabolic solution continuously connects with the overnutrition solution, $\mu^* = d(\alpha_3 g - 1)$ and $m_{\rm cat}^* = 0$. In addition, one can easily find that $\mu^*$ becomes negative below $\alpha_3g < 1-\alpha_1$, which indicates the transition to the starvation state.

\section{Asymptotic solutions for low nutrient supply}
\label{sec:asymptotic}

In this section we consider the situation where $\mu^* \ll \alpha_3g \ll 1$ and $\mu^* \ll d$. We show that, for both the uniform distribution with $r=1$ and the power-law distribution, 
the asymptotic behavior of $\mu^*$ as a function of $\alpha_3 g$ exhibits the exponential form $\exp(-C/\alpha_3 g)$, where $C$ is a positive constant.

The average abundance of chemical species belonging to $\mathcal{G}_n$ is given by $m_n^* = (d_n g_n + m_{\rm in}^* \mcat^*)X(\mu^*+d_n, \mcat^*)$ in the main text, where $X(a,b) = \sum_{v}\pout/(a+\frac{v}{c}b)$. Hence, if $\mu^* \ll d$, fixed-point values of the average abundances of unpenetrable metabolic products, penetrable metabolic products, and nutrients are given by
\begin{align}
m_1^* &= X(\mu^*,\mcat^*)m_{\rm in}^*\mcat^*, \label{eq:m1asympt} \\
m_2^* & \simeq X(d,\mcat^*)m_{\rm in}^*\mcat^*, \label{eq:m2asympt} \\
m_3^* & \simeq X(d,\mcat^*)(dg + m_{\rm in}^*\mcat^*), \label{eq:m3asympt}
\end{align}
respectively. Substituting these into the definition of $\mu$ and using Eq.~{\bf 4} in the main text, we obtain the relation
\begin{align}
X(\mu^*,\mcat^*) \simeq \frac{1-\alpha_3  g}{\alpha_1 m_{\rm in}^*\mcat^*}.
\label{eq:xmumcat}
\end{align}
Because the uniform distribution with $r=1$ and the power-law distribution with $\beta=3$ with dense connectivity, $c\gg1$, are written as $p(v/c)=1/2$ with the interval $[0,2]$ and $p(v/c)=2/(1+v/c)^3$ with the interval $[0,\infty)$, respectively, the function $X(a,b)$ can be evaluated as
\begin{align}
X(a,b) &= \sum_v \pout/(a+\frac{v}{c}b) \nonumber \\
&=
\begin{cases}
&\int_0^2 dv' 1/2(a+v'b)~~~~~~~~~~~~(\text{uniform}) \\
&\int_0^\infty 2/(1+v')^3(a+v'b)~~~~(\text{power-law})
\end{cases} \nonumber \\
&=
\begin{cases}
&\frac{1}{2b}\log\frac{a+2b}{a}~~~~~~~~~~~~~~~~~~~~~~(\text{uniform}) \\
&\frac{1}{(a-b)^2}\qty(\frac{2b^2}{a-b}\log\frac{b}{a}+a-3b)~(\text{power-law}),
\end{cases}
\end{align}
where $v'=v/c$. One finds that the value of $X(a,b)$ logarithmically diverges as $a\to+0$.
Thus, when $\mu^* \ll \alpha_3g \ll 1$, the relation given by Eq.~\refbf{eq:xmumcat} leads to
\begin{align}
\mu^* =
\begin{cases}
&2\mcat^*\exp\qty(-2\frac{1-\alpha_3g}{\alpha_1m_{\rm in}^*})~~~~~(\text{uniform}) \\
&\mcat^*\exp\qty(-\frac{3}{2}-\frac{1-\alpha_3g}{2\alpha_1m_{\rm in}^*})~~(\text{power-law}).
\end{cases}
\label{eq:mu_relation}
\end{align}

Next, we derive $\mcat^*$ and $m_{\rm in}^*$ as functions of $\alpha_3 g$. Using Eqs. \refbf{eq:m1asympt}, \refbf{eq:m2asympt}, \refbf{eq:xmumcat}, and Eq.~{\bf 7} in the main text, we obtain
\begin{align}
m_{\rm in}^* = \frac{(\alpha_1+\alpha_2)\mcat^*-1+\alpha_3g}{\mcat^*\alpha_2X(d,\mcat^*)}.
\label{eq:min1}
\end{align}
Using Eqs.~\refbf{eq:m1asympt}-\refbf{eq:m3asympt} and Eq. {\bf 4} in the main text, we obtain
\begin{align}
m_{\rm in}^* = \frac{(1-dX(d,\mcat^*))\alpha_3g}{(1-\alpha_1)\mcat^*X(d,\mcat^*)}.
\label{eq:min2}
\end{align}
From Eqs.~\refbf{eq:min1} and \refbf{eq:min2}, we can derive the equation for $\mcat^*$ as
\begin{align}
(\alpha_1+\alpha_2)\mcat^*-1+\alpha_3g = \frac{\alpha_2(1-dX(d,\mcat^*))\alpha_3 g}{1-\alpha_1}.
\end{align}
If $\alpha_3g \ll 1$, we find from this equation that $\mcat^*$ can be expanded in powers of $\alpha_3g$ as
\begin{align}
\mcat^* = \frac{1+A_1\alpha_3g+A_2(\alpha_3g)^2+\cdots}{\alpha_1+\alpha_2}.
\label{eq:mcat}
\end{align}
The coefficients are given by
\begin{align}
A_1 &= -1 + \frac{\alpha_2}{1-\alpha_1}(1-dX(d,\frac{1}{\alpha_1+\alpha_2})), \\
A_2 &= -\frac{\alpha_2dA_1}{(1-\alpha_1)(\alpha_1+\alpha_2)}\left.\pdv{X(d,\mcat)}{\mcat}\right|_{\mcat = 1/(\alpha_1+\alpha_2)}.
\end{align}
Substituting Eqs.~\refbf{eq:min2}, and \refbf{eq:mcat} into Eq.~\refbf{eq:mu_relation}, we obtain the asymptotic behavior of $\mu^*$ as a function of $\alpha_3g$ in the regime $\mu^* \ll \alpha_3g \ll 1$ and $\mu^* \ll d$. 
Since $m_{\rm in}^\ast$ is proportional to $\alpha_3 g$, $\mu^\ast$ exhibits an essential singularity as $\alpha_3 g \to 0$, thereby justifying the assumption $\mu^\ast \ll \alpha_3 g$.






\begin{figure}[t]
\centering
\includegraphics[width=130mm]{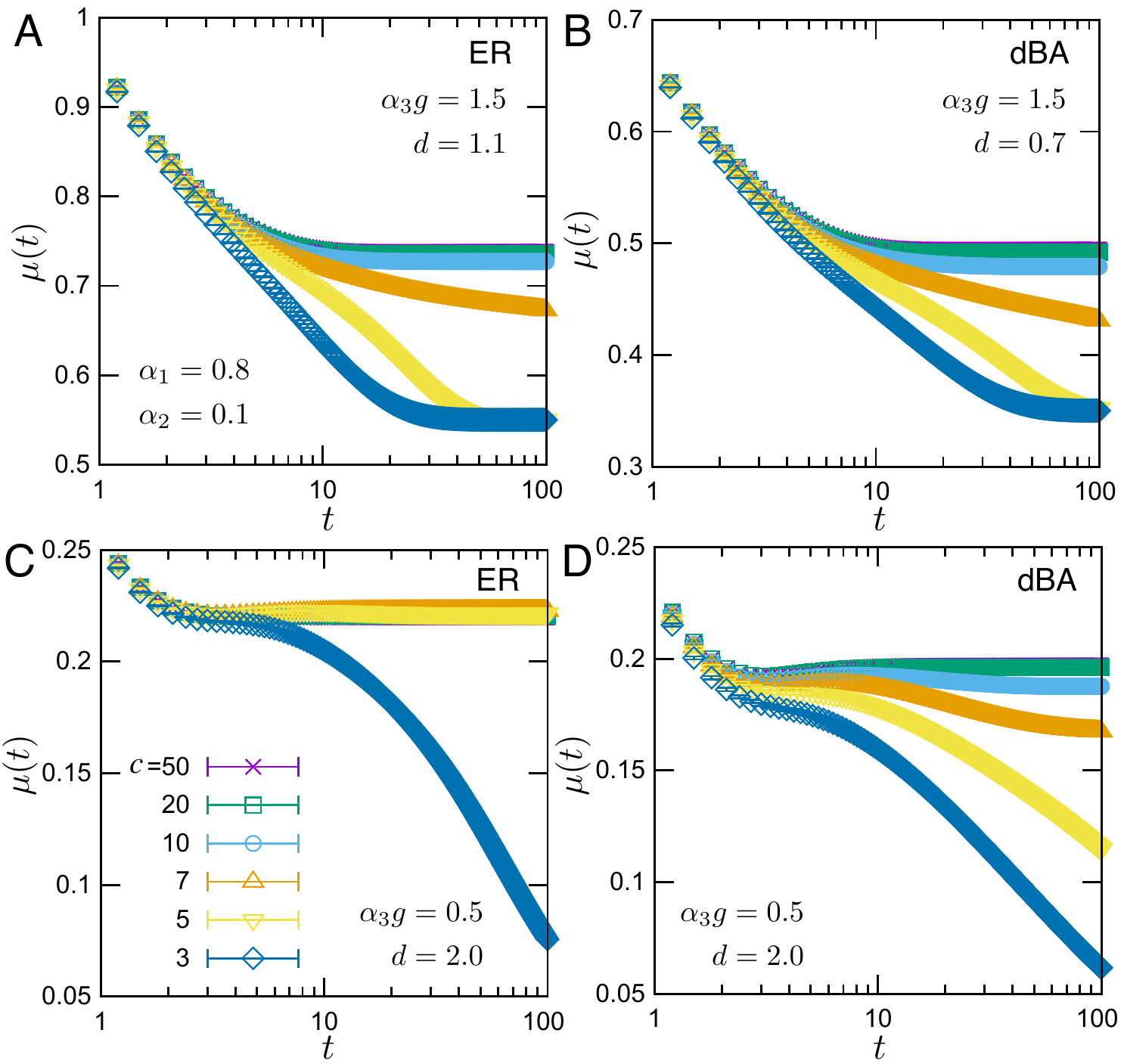}
\caption{
Time dependence of growth rate $\mu(t)$ at (A) $(\alpha_3g,d) = (1.5,1.1)$ and (C) $(\alpha_3g,d) = (0.5,2.0)$ for Erd\H{o}s-R\'{e}nyi networks and (B) $(\alpha_3g,d) = (1.5,0.7)$ and (D) $(\alpha_3g,d) = (0.5,2.0)$ for directed Barab\'{a}si-Albert networks with finite connectivity $c = 50, 20, 10, 7, 5$ and 3. Simulations are performed with parameters $N=1.0 \times 10^5$, $\alpha_1 = 0.8$ and $\alpha_2 = 0.1$ and averaged over 100 network realizations. One finds that $t=100$ is sufficient to reach the fixed point for networks with $ c\ge 10$.
}
\label{fig:dynamics}
\end{figure}

\begin{figure}[t]
\centering
\includegraphics[width=170mm]{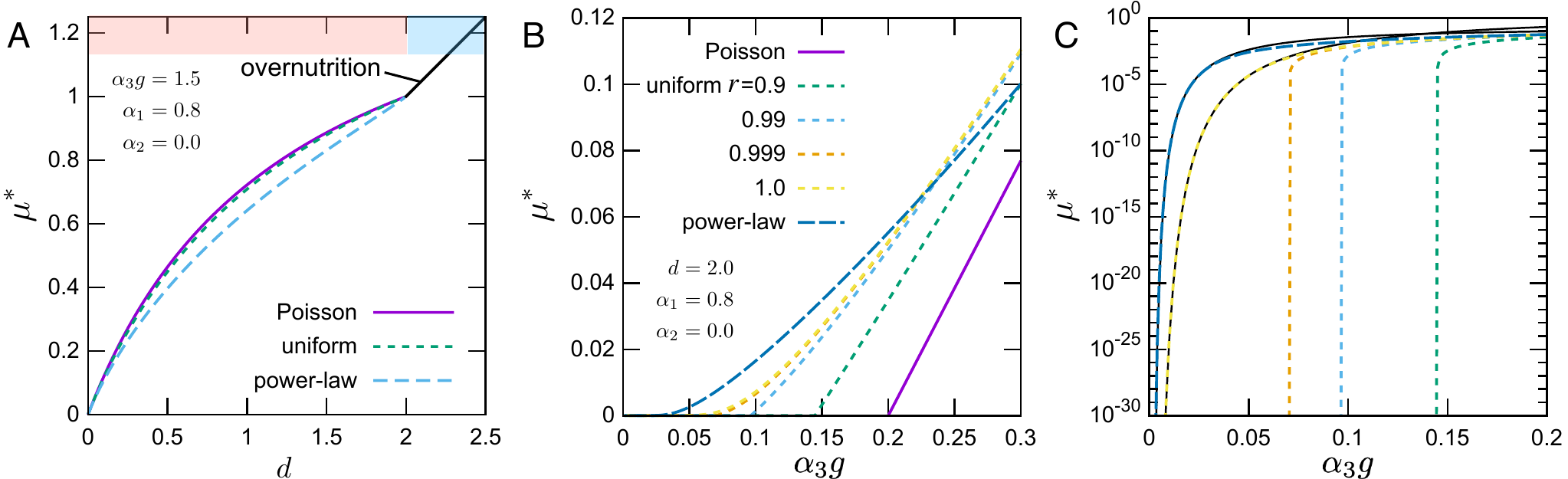}
\caption{
Behaviors of the growth rate without the penetrable metabolic products $\mathcal{G}_2$, i.e., $\alpha_2 = 0$. (A) Dependence on permeability $d$ of the growth rate $\mu^*$ in steady state for $\alpha_3g=1.5$, obtained from the DMFT. In the metabolic state $d<d_{\rm c} = 1/(\alpha_3 g - 1) = 2$, different colors represent different network topologies: results for Poisson, uniform ($r=1$), and power-law degree distributions (exponent $\beta = 3$) are shown from top to bottom. The black line for $d>d_{\rm c}$ represents the non-metabolic state. (B) Dependence on nutrient supply $\alpha_3g$ of the growth rate $\mu^*$ obtained from the DMFT and (C) its semi-log plot for $d=2.0$. Different colors represent different network topologies: results for Poisson, uniform ($r=0.9,0.99,0.999,1$), and power-law degree distributions (exponent $\beta = 3$) are shown. The black solid curves in panel (C) represent asymptotic solutions for $\alpha_3g \ll 1$. $\alpha_1 = 0.8$ is used in all panels.
}
\label{fig:dynamics}
\end{figure}

\end{document}